\documentclass[twocolumn,showpacs,prb,a4paper,floatfix]{revtex4-1}
\usepackage{amssymb}
\usepackage{graphicx}
\usepackage{amsmath}
\usepackage{graphicx}
\usepackage{dcolumn}
\usepackage{bm}
\usepackage{color}

\newcommand{\bi}{\begin{itemize}}
\newcommand{\ei}{\end{itemize}}
\newcommand{\be}{\begin{equation}}
\newcommand{\ee}{\end{equation}}
\newcommand{\bea}{\begin{eqnarray}}
\newcommand{\eea}{\end{eqnarray}}

\begin{document}

\title{Landau Level Spectrum of ABA- and ABC-stacked Trilayer Graphene}
\author{Shengjun Yuan$^{1}$, Rafael Rold\'{a}n$^{1,2}$, and Mikhail I.
Katsnelson$^1$}
\date{\today }
\affiliation{\centerline{$^1$Institute for Molecules and Materials, Radboud University of Nijmegen, NL-6525AJ Nijmegen, The Netherlands}\\
\centerline{$^2$Instituto de Ciencia de Materiales de Madrid, CSIC, Cantoblanco E28049 Madrid, Spain}}

\begin{abstract}
We study the Landau level spectrum of ABA- and ABC-stacked trilayer
graphene. We derive analytic low energy expressions for the spectrum, the
validity of which is confirmed by comparison to a $\pi $-band tight-binding
calculation of the density of states on the honeycomb lattice. We further
study the effect of a perpendicular electric field on the spectrum, where a
zero-energy plateau appears for ABC stacking order, due to the opening of a
gap at the Dirac point, while the ABA-stacked trilayer graphene remains
metallic. We discuss our results in the context of recent electronic
transport experiments. Furthermore, we argue that the expressions obtained
can be useful in the analysis of future measurements of cyclotron resonance
of electrons and holes in trilayer graphene.
\end{abstract}

\maketitle

\section{Introduction}

Recent experimental realizations of graphene trilayers\cite%
{TJ11,BL11,LH11,KR11,ZZ11,JS11} (TLG) have opened the possibility of
exploring their intriguing electronic properties, which depend dramatically
on the stacking sequence of the graphene layers.\cite{GCP06} The low energy
band structure for ABA-stacked TLG consists of one massless and two massive
subbands, similar to the spectrum of one single layer (SLG) and one bilayer
graphenes (BLG), while ABC trilayer presents approximately cubic bands.\cite%
{KM09} Interestingly, when the TLG is subjected to a perpendicular electric
field, a gap can be opened for ABC samples,\cite%
{Avetisyan2009,Avetisyan2010,Zhang2010,BL11,LH11} similarly to bilayer
graphene,\cite{CC07} whereas ABA TLG remains metallic with a tunable band
overlap.\cite{CT09}

When a strong magnetic field is applied perpendicular to the TLG planes, the
band structure is quantized into Landau levels (LLs). The number of graphene
layers as well as their relative orientation (stacking sequence) determine
the features of the quantum Hall effect (QHE) in this material, where the
Hall conductivity presents plateaus at\cite{MF06,MM08} 
\begin{equation}
\sigma_{xy}=\pm\frac{ge^2}{h}\left(n+\frac{N}{2}\right),
\end{equation}
where $N=3$ is the number of layers, $n$ is the LL index, $g=4$ is the LL
degeneracy due to spin and valley degrees of freedom, $-e$ is the electron
charge and $h$ is the Planck's constant. In particular, the plateau
structure in $\sigma_{xy}$ of TLG has been shown to be strongly dependent on
the stacking sequence.\cite{BL11}

In this paper we study the LL quantization of TLG. We obtain analytical
expressions for the LL spectrum of TLG with ABA or ABC stacking order. The
range of applicability of the analytical results is studied by a comparison
to the density of states (DOS) obtained from a numerical solution of the
time-dependent Schr\"{o}dinger equation within the framework of a
tight-binding model on the honeycomb lattice.\cite{HR00,YRK10,YRK10b} We
further study the effect of a perpendicular electric field in the LL
spectrum, finding that a zero-energy plateau develops in the Hall
conductivity only for ABC-stacked graphene, while ABA-stacked graphene
remains ungapped.

The paper is organized as follows. In Sec. \ref{Sec:LLspectrum} we obtain
analytically the low energy LL spectrum of TLG. The analytic expressions of
Sec. \ref{Sec:LLspectrum} are compared to the DOS numerically obtained from
a full tight-binding calculation in the honeycomb lattice in Sec. \ref%
{Sec:Num}. Our main conclusions are summarized in Sec. \ref{Sec:Concl}.

\section{Analytic derivation of the Landau level spectrum}

\label{Sec:LLspectrum}

\begin{figure}[t]
\begin{center}
\mbox{
\includegraphics[width=4cm]{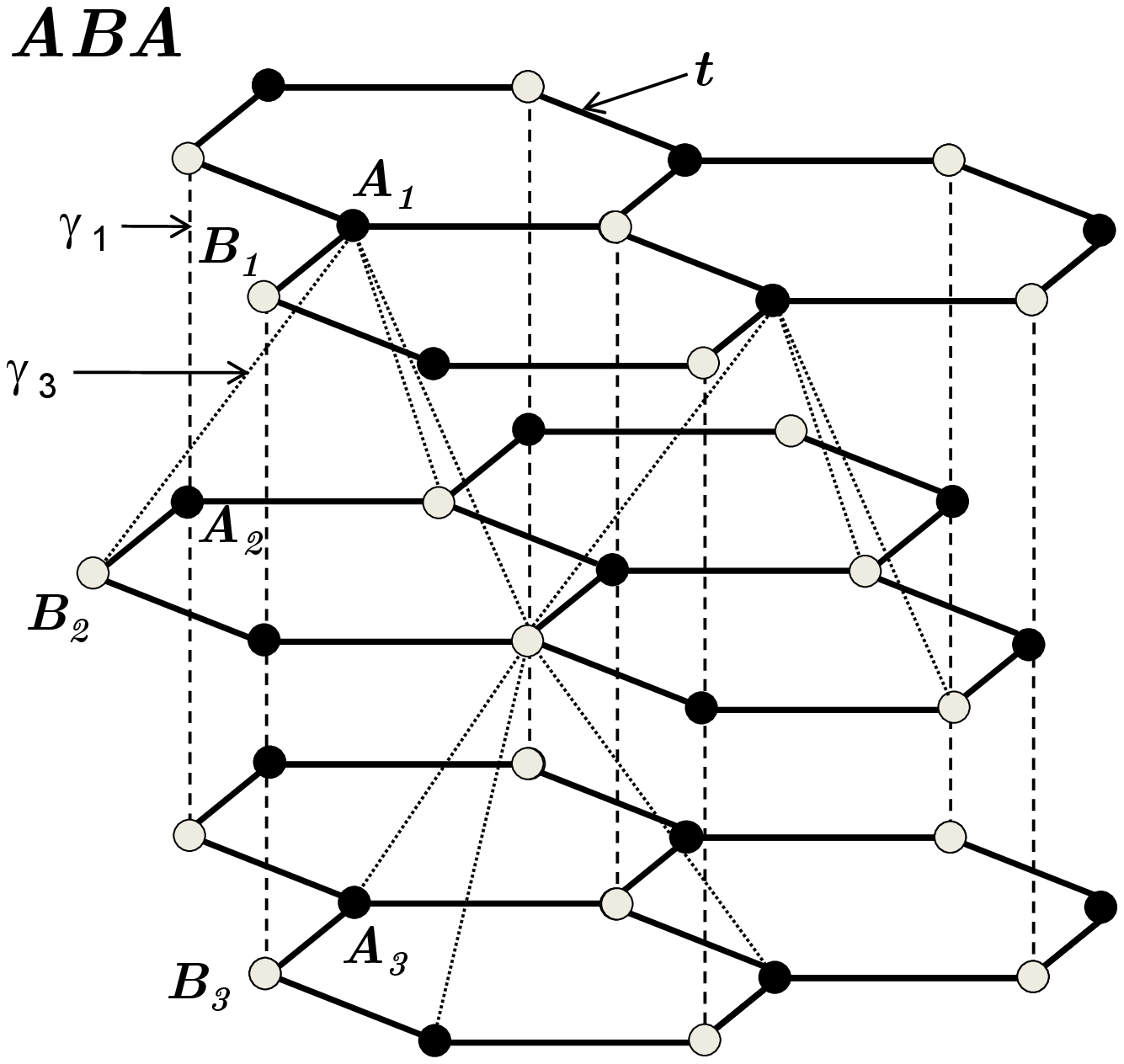}
\includegraphics[width=4cm]{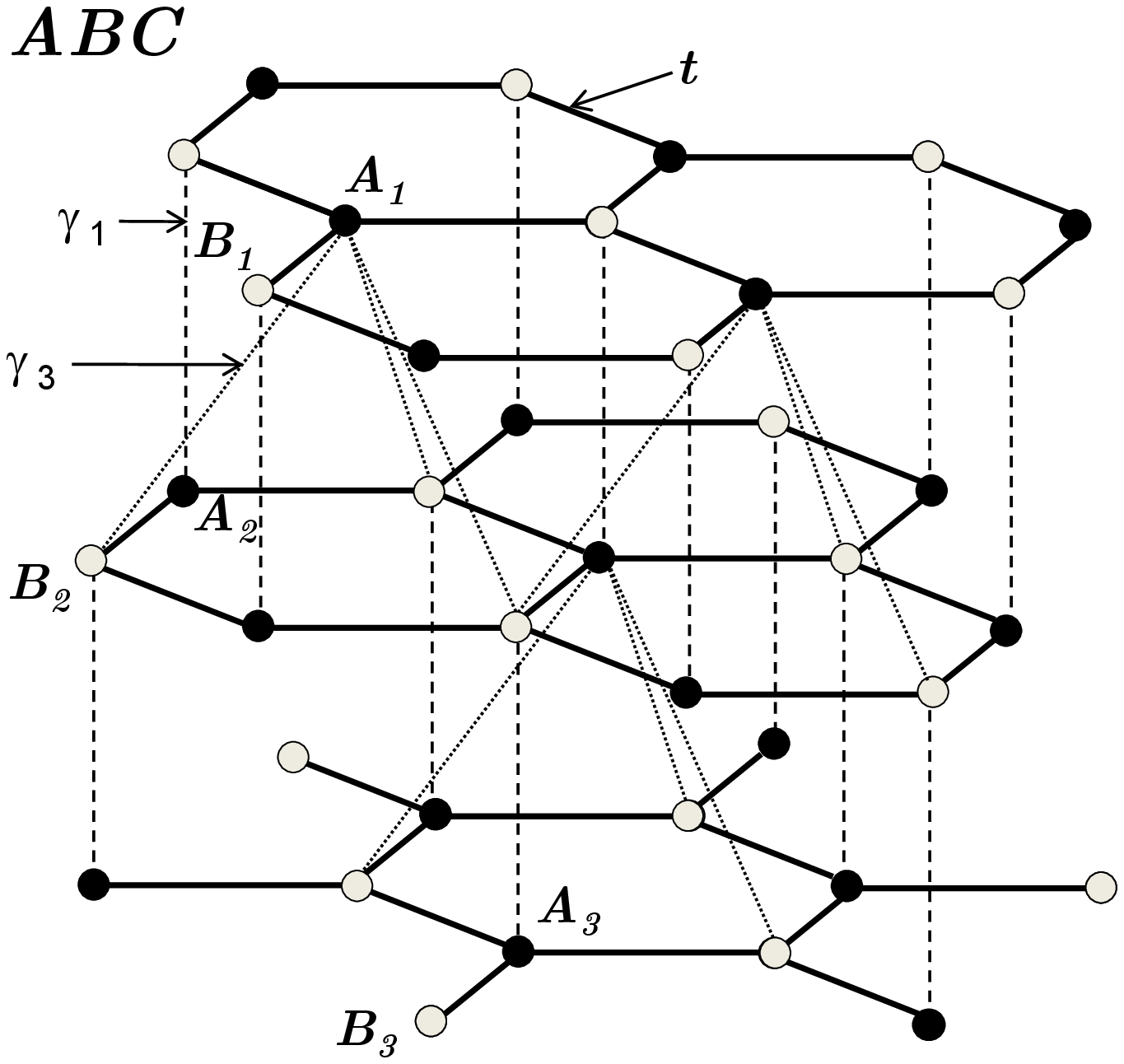}
}
\end{center}
\caption{Atomic structure of ABA- and ABC-stacked trilayer graphene. The
intra-layer $t$ and inter-layer $\protect\gamma_1$ and $\protect\gamma_3$
hopping amplitudes are schematically shown in the figure.}
\label{Fig:Stacking}
\end{figure}

\begin{figure}[t]
\begin{center}
\mbox{
\includegraphics[width=4.2cm]{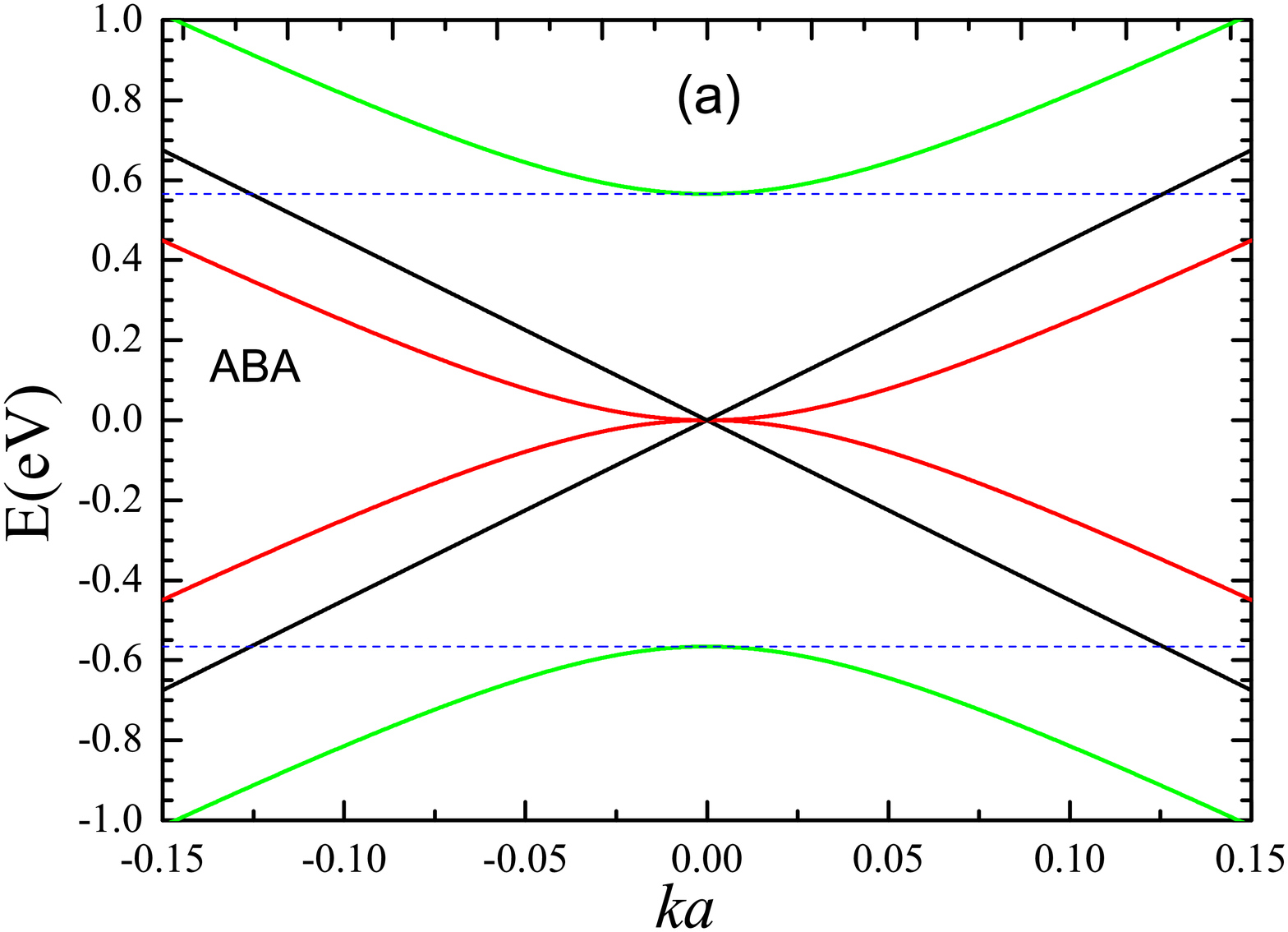}
\includegraphics[width=4.2cm]{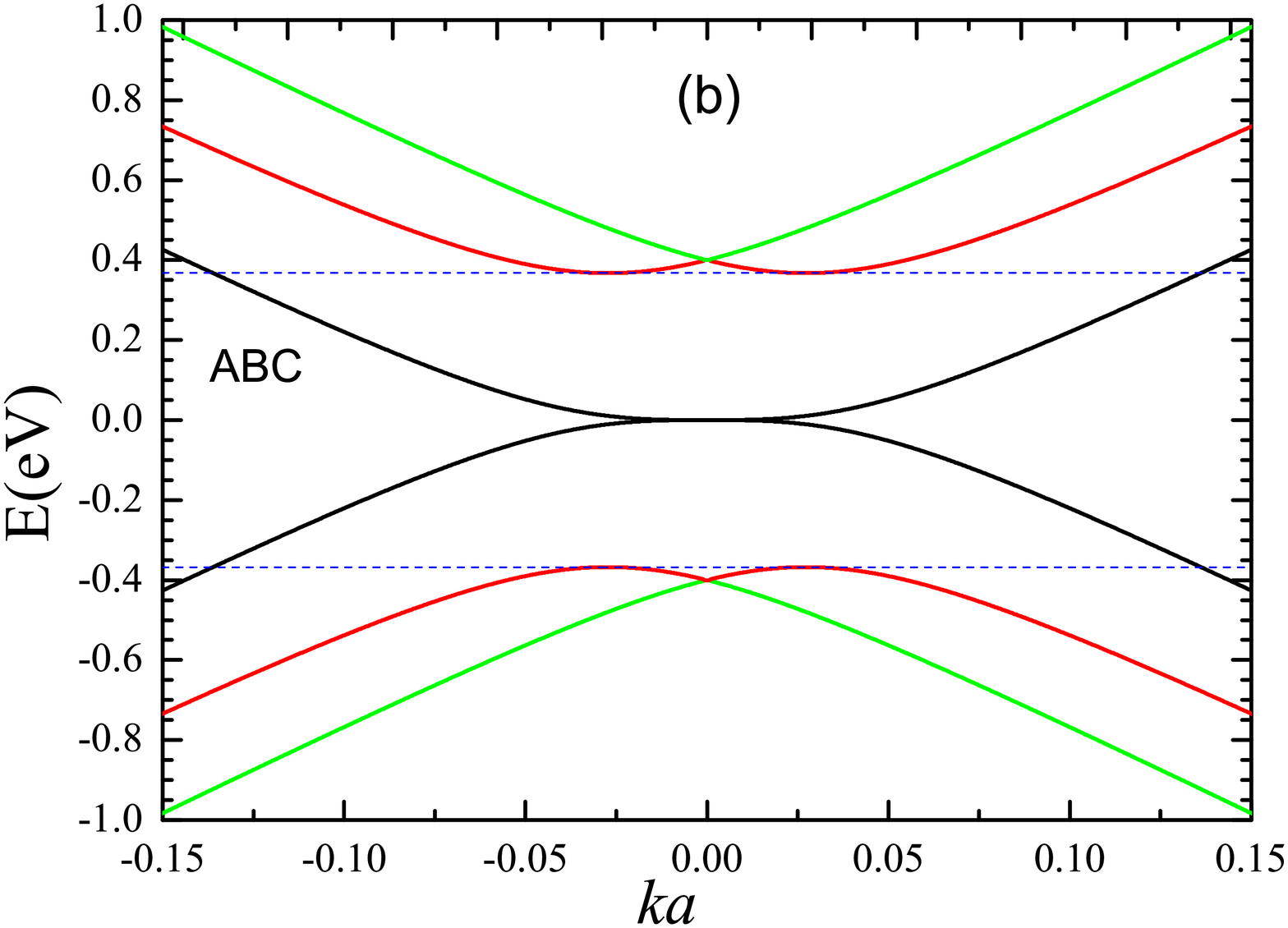}
}
\end{center}
\caption{(Color online) Low energy band structure of ABA- and ABC-stacked
trilayer graphene around the $K$ point. We have used the tight-binding
parameters $t=3$~eV and $\protect\gamma _{1}=0.4$~eV. The red dashed lines
are a guide to the eye that mark, for the used parameters $t$ and $\protect%
\gamma _{1}$, the position of the bottom (top) of the upper (lower) bands.
The analytic expressions of these bands are given in Appendix A.}
\label{Fig:Bands}
\end{figure}

In nature there are two known forms of stable stacking sequence in TLG,
namely ABA (Bernal) and ABC (rhombohedral) stacking. The difference between
ABA and ABC stacking, schematically shown in Fig. \ref{Fig:Stacking}, is
that the third layer is rotated with respect to the second layer by $%
-120^{\circ }$ (so that it will be exactly under the first layer) in ABA
stacking, while it is rotated by $+120^{\circ }$ in ABC stacking.\cite%
{GCP06,K10} In a basis with components of $\psi _{A_{1}},$ $\psi _{B_{1}},$ $%
\psi _{A_{2}},$ $\psi _{B_{2}},$ $\psi _{A_{3}},$ $\psi _{B_{3}}$, where $%
\psi _{A_{i}}$ ($\psi _{B_{i}}$) are the envelope functions associated with
the probability amplitudes of the wave functions on the sublattice A (B) of
the $i$th layer ($i=1,2,3$), the effective low energy Hamiltonian of
ABA-stacked TLG around the $K$ point is\cite{GCP06}%
\begin{equation}
H_{\mathbf{p}}=\left( 
\begin{array}{cccccc}
0 & v_{\mathrm{F}}p_{-} & 0 & 0 & 0 & 0 \\ 
v_{\mathrm{F}}p_{+} & 0 & \gamma _{1} & 0 & 0 & 0 \\ 
0 & \gamma _{1} & 0 & v_{\mathrm{F}}p_{-} & 0 & \gamma _{1} \\ 
0 & 0 & v_{\mathrm{F}}p_{+} & 0 & 0 & 0 \\ 
0 & 0 & 0 & 0 & 0 & v_{\mathrm{F}}p_{-} \\ 
0 & 0 & \gamma _{1} & 0 & v_{\mathrm{F}}p_{+} & 0%
\end{array}%
\right) ,  \label{Eq:HamABA0}
\end{equation}%
where $p_{\pm }=p_{x}\pm ip_{y}$, with $\mathbf{p}=\left( p_{x},p_{y}\right) 
$ the two-dimensional momentum operator, and $v_{F}=3at/2$ the Fermi
velocity of the monolayer graphene, in terms of the in-plane nearest
neighbor hopping $t\approx 3$~eV and the carbon-carbon distance $a\approx
1.42$~\AA ~(from now on we use units such that $\hbar \equiv 1\equiv c$).
For the moment, we only include the inter-layer hopping $\gamma _{1}\approx
0.4$~eV in Eq. (\ref{Eq:HamABA0}). The effective Hamiltonian for $K^{\prime
} $ is obtained by exchanging $p_{+}$ and $p_{-}$. The effect of far-distant
hopping such as $\gamma _{3}$ will be discussed in Appendix \ref{App:gamma3}%
. the Hamiltonian (\ref{Eq:HamABA0}) leads to a combination of two linear
SLG-like bands [black lines in Fig. \ref{Fig:Bands}(a)] and four massive
BLG-like bands [red and green lines in Fig. \ref{Fig:Bands}(a)].

\begin{figure*}[t]
\begin{center}
\mbox{
\includegraphics[width=8cm]{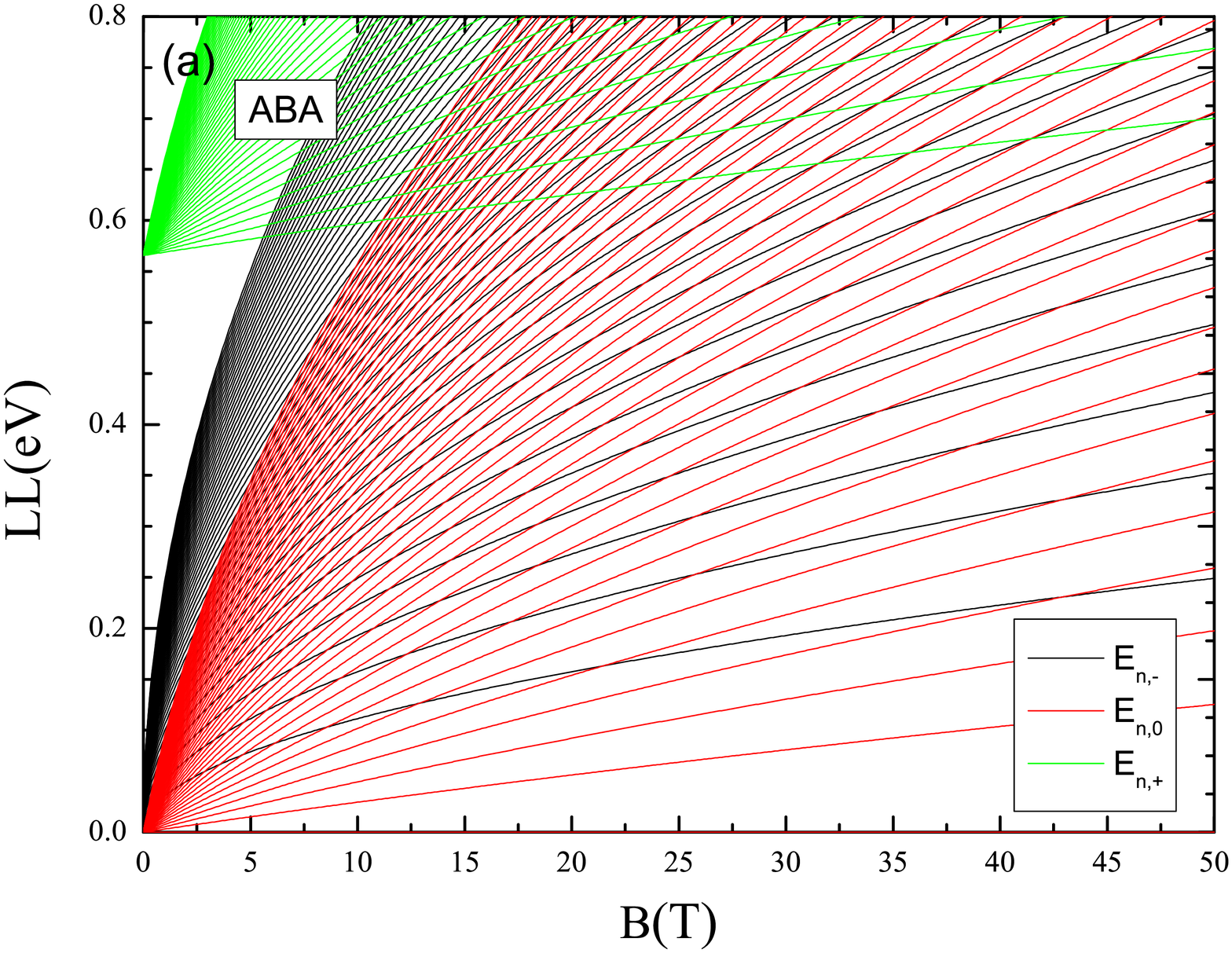}
\includegraphics[width=8cm]{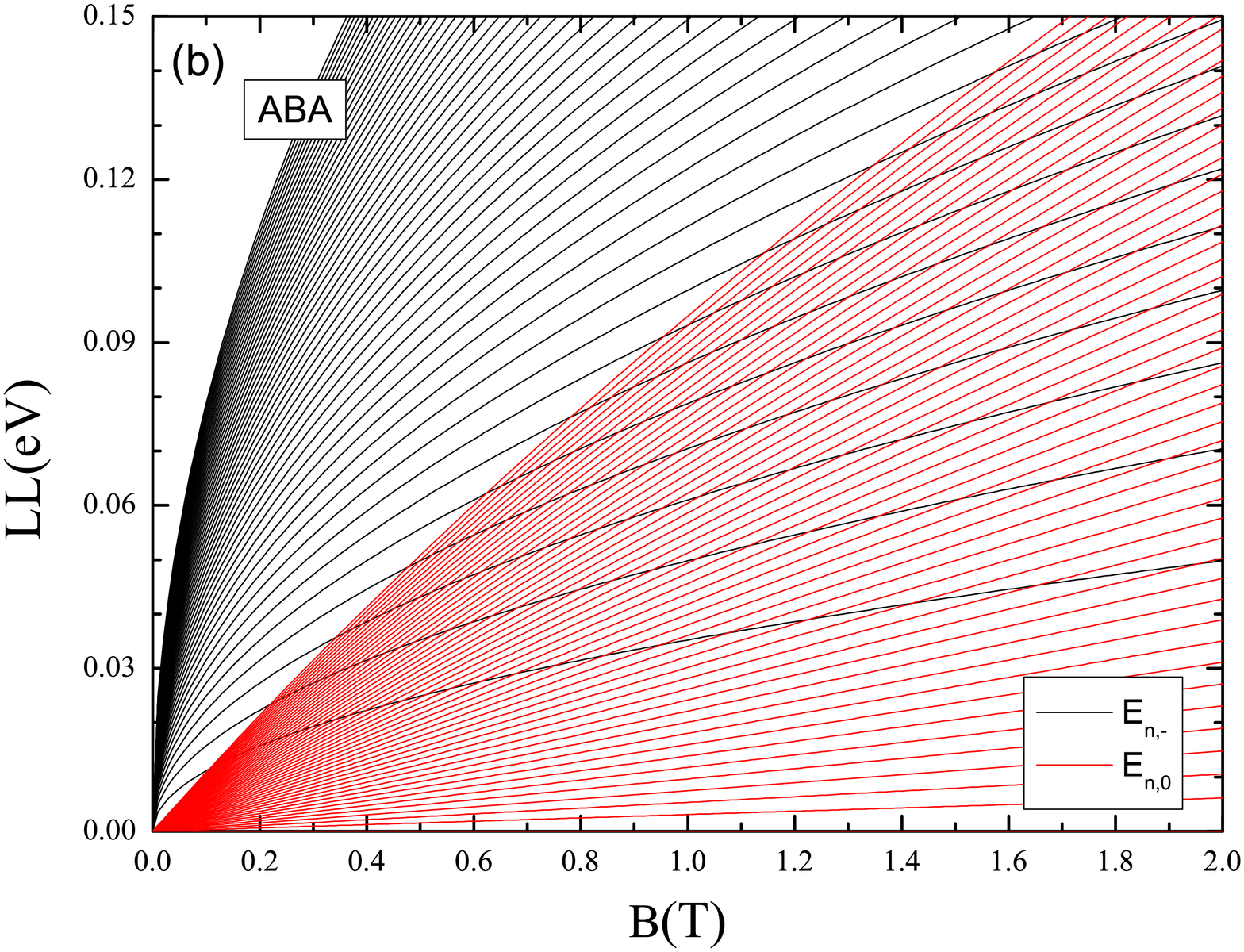}
} 
\mbox{
\includegraphics[width=8cm]{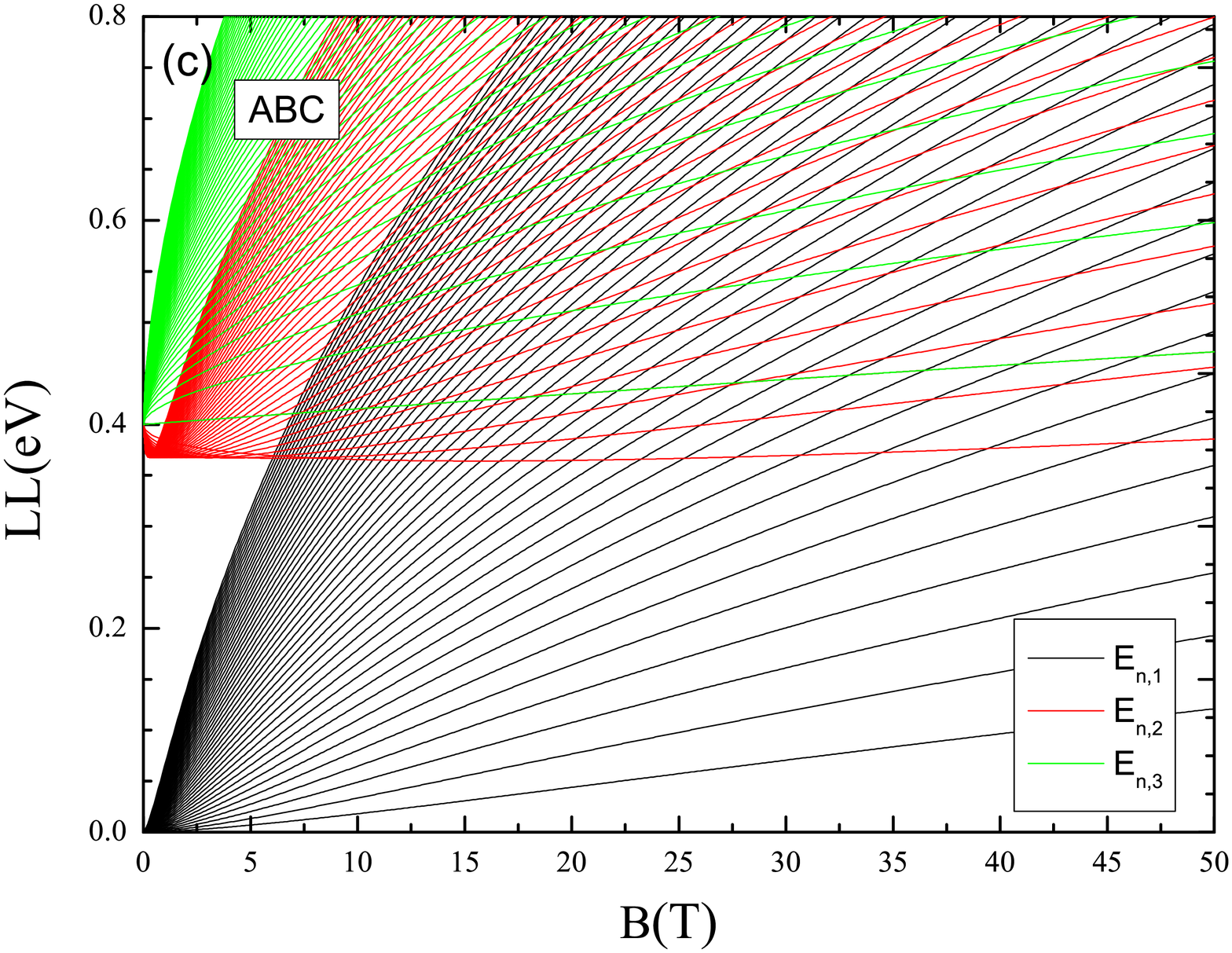}
\includegraphics[width=8cm]{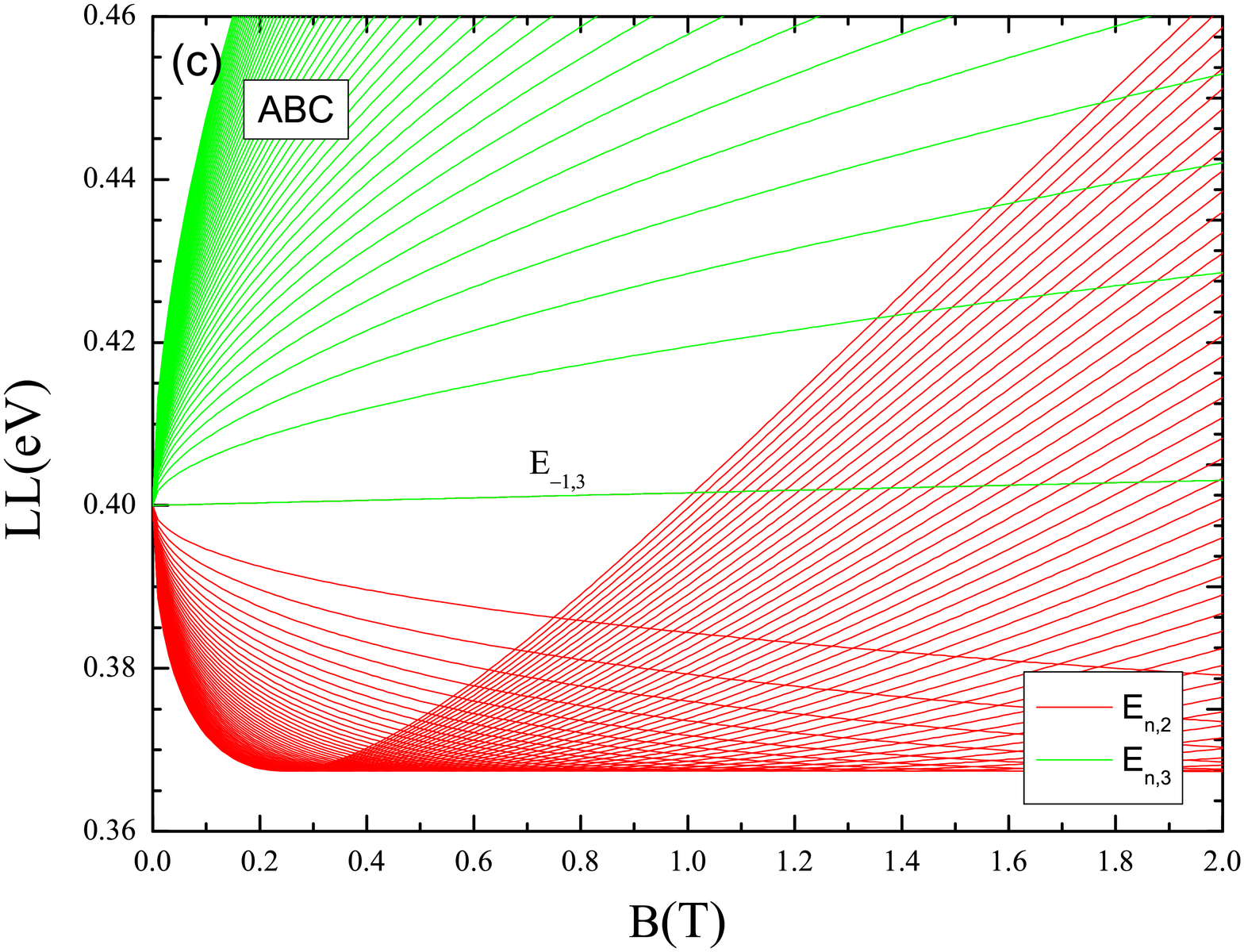}
}
\end{center}
\caption{(Color online) Three band structures in the Landau level spectrum
of the ABA- and ABC-stacked trilayer graphene. We have used Eq. (\protect\ref%
{Eq:DispABA-BLG})-(\protect\ref{Eq:DispABA-SLG}) for ABA stacking, and Eq. (%
\protect\ref{Eq:LLsABC}) for ABC stacking. Only the first 50 Landau levels
in each band are presented.}
\label{Fig:LLs}
\end{figure*}

In the presence of an external perpendicular magnetic field,\cite{G11} the
canonical momentum $\mathbf{p}$ must be replaced by the gauge-invariant
kinetic momentum ${\mathbf{p}}\rightarrow {\boldsymbol{\Pi }}=\mathbf{p}+e%
\mathbf{A}(\mathbf{r})$ where $\mathbf{A}(\mathbf{r})$ is the vector
potential, and which obey the commutation relation $[\Pi _{x},\Pi
_{y}]=-i/l_{B}^{2}$, where $l_{B}=1/\sqrt{eB}$ is the magnetic length.
Therefore, this allows to introduce the ladder operators $\hat{a}=(l_{B}/%
\sqrt{2})\Pi _{-}$ and $\hat{a}^{\dagger }=(l_{B}/\sqrt{2})\Pi _{+}$, where $%
\Pi _{\pm }=\Pi _{x}\pm i\Pi _{y}$, and which obey the commutation relation $%
[\hat{a},\hat{a}^{\dagger }]=1$. As in the usual one-dimensional harmonic
oscillator, 
\begin{equation*}
\hat{a}\left\vert n\right\rangle =\sqrt{n}\left\vert n-1\right\rangle ,\hat{a%
}^{\dagger }\left\vert n\right\rangle =\sqrt{n+1}\left\vert n+1\right\rangle
,
\end{equation*}%
where $|n\rangle $ is an eigenstate of the usual number operator $\hat{a}%
^{\dagger }\hat{a}|n\rangle =n|n\rangle $, with $n\geq 0$ an integer. Then,
the Hamiltonian can be expressed in terms of $\hat{a}$ and $\hat{a}^{\dagger
}$ as 
\begin{equation}
\mathcal{H}=\left( 
\begin{array}{cccccc}
0 & \Delta _{B}\hat{a} & 0 & 0 & 0 & 0 \\ 
\Delta _{B}\hat{a}^{\dagger } & 0 & \gamma _{1} & 0 & 0 & 0 \\ 
0 & \gamma _{1} & 0 & \Delta _{B}\hat{a} & 0 & \gamma _{1} \\ 
0 & 0 & \Delta _{B}\hat{a}^{\dagger } & 0 & 0 & 0 \\ 
0 & 0 & 0 & 0 & 0 & \Delta _{B}\hat{a} \\ 
0 & 0 & \gamma _{1} & 0 & \Delta _{B}\hat{a}^{\dagger } & 0%
\end{array}%
\right) ,  \label{Eq:Ha}
\end{equation}%
where $\Delta _{B}$ is the magnetic energy defined by $\Delta _{B}=\sqrt{2}%
v_{\mathrm{F}}/l_{B}$. Therefore the six-components eigenstates of $\mathcal{%
H}$ can be reconstructed as $\psi ={\Large [}c_{A_{1}}\varphi _{n-1,k}$, $%
c_{B_{1}}\varphi _{n,k}$, $c_{A_{2}}\varphi _{n,k}$, $c_{B_{2}}\varphi
_{n+1,k}$, $c_{A_{3}}\varphi _{n-1,k}$, $c_{B_{3}}\varphi _{n,k}{\Large ]}%
^{T},$ where $c_{A_{i}}(c_{B_{i}})$ are amplitudes. If we choose the Landau
gauge $\mathbf{A}(\mathbf{r})=(0,Bx)$, then the wave function of the $n$th
LL $\varphi _{n,k}(x,y)$ is given by \cite{KA08} 
\begin{equation}
\varphi _{n,k}(x,y)=i^{n}\left( \frac{1}{2^{n}n!\sqrt{\pi }{l_{B}}}\right)
^{1/2}e^{iky}e^{-z^{2}/2}H_{n}\left( z\right) ,
\end{equation}%
where $z=(x-kl_{B}^{2})/l_{B}$, $H_{n}\left( z\right) $ is the Hermite
polynomial, and $\varphi _{n,k}\equiv 0$ for $n<0$. Then, the Hamiltonian
matrix in the basis of $\psi $ is 
\begin{equation}
\left( 
\begin{array}{cccccc}
0 & \Delta _{B}C_{1} & 0 & 0 & 0 & 0 \\ 
\Delta _{B}C_{1} & 0 & \gamma _{1} & 0 & 0 & 0 \\ 
0 & \gamma _{1} & 0 & \Delta _{B}C_{2} & 0 & \gamma _{1} \\ 
0 & 0 & \Delta _{B}C_{2} & 0 & 0 & 0 \\ 
0 & 0 & 0 & 0 & 0 & \Delta _{B}C_{1} \\ 
0 & 0 & \gamma _{1} & 0 & \Delta _{B}C_{1} & 0%
\end{array}%
\right) ,  \label{Eq:HamiltonianABA}
\end{equation}%
with $C_{1}=\sqrt{n}$ and $C_{2}=\sqrt{n+1}$. Eq. (\ref{Eq:HamiltonianABA})
has six eigenvalues, which can be easily calculated:%
\begin{eqnarray}  \label{Eq:LLsABA}
E_{n,s} &=&\pm \frac{1}{\sqrt{2}}{\Large [}2\gamma _{1}^{2}+\left(
2n+1\right) \Delta _{B}^{2}  \notag  \label{BandTLGABA} \\
&&+s\sqrt{4\gamma _{1}^{4}+4\left( 2n+1\right) \gamma _{1}^{2}\Delta
_{B}^{2}+\Delta _{B}^{4}}{\Large ]}^{1/2},  \label{Eq:DispABA-BLG} \\
E_{n,0} &=&\pm \Delta _{B}\sqrt{n},  \label{Eq:DispABA-SLG}
\end{eqnarray}%
with $s=\pm 1$ and $n\geq 0$. The eigenstates corresponding to above LLs are
given in Appendix B. Notice that Eq. (\ref{Eq:DispABA-BLG}) coincides (apart
from a numerical factor $\sqrt{2}$ in front of $\gamma _{1}$) with the LL
spectrum of a bilayer graphene,\cite{PPV07} whereas the Eq. (\ref%
{Eq:DispABA-SLG}) corresponds to the LL spectrum of a single layer graphene.
This is expected since the low energy band structure of ABA TLG consists of
two massless SLG-like bands and four massive BLG-like bands, as it has been
discussed above. In Fig. \ref{Fig:LLs}(a) we show the LL spectrum Eq. (\ref%
{Eq:LLsABA})-(\ref{Eq:DispABA-SLG}) for ABA TLG obtained for the first 50
LLs of each band (we only show the states with positive energy). As in the
zero magnetic field case, there are two sets of BLG-like LLs which disperse
roughly linearly with $B$ (the LLs plotted in red and green color), whereas
the linearly in $\mathbf{k}$ dispersing SLG-like band leads to a set of $%
\sqrt{B}$-like LLs (plotted in black) [see Fig. \ref{Fig:LLs}(b) for a zoom
of the low energy and low magnetic field region of Fig. \ref{Fig:LLs} (a)].
Furthermore, a set of LL crossings occur due to the massless and massive
characters of the subbands, as it has been observed experimentally.\cite%
{TJ11} Notice that the Landau levels in the low energy part of the spectrum
have only $E_{n,-}$ character [see Fig. \ref{Fig:LLs}(a) and (b)], unless
the magnetic field is very strong. For example, the third low energy Landau
level belongs to the set of LLs $E_{n,0}$ when $B\gtrsim 45$~T. On the other
hand, the $E_{n,+}$ LLs only appear at an energy $\left\vert E\right\vert
\geq \left\vert E_{0,+}\right\vert =\sqrt{2\gamma _{1}^{2}+\Delta _{B}^{2}}$%
. In the limit $n\Delta _{B}^{2}\ll \gamma _{1}^{2}$, the BLG-like bands Eq.
(\ref{Eq:DispABA-BLG}) can be simplified to 
\begin{equation}
E_{n,-}\approx \pm \frac{v_{\mathrm{F}}^{2}}{l_{B}^{2}\gamma _{1}}\sqrt{%
2n\left( n+1\right) },  \label{Eq:LLsABAapprox}
\end{equation}%
which is similar to the commonly used expression for the low energy spectrum
of BLG in a weak magnetic field.\cite{MF06}

Whereas some of the results for the LL spectrum of ABA trilayer graphene has
been discussed before,\cite{KM11} much less effort has been put on
understanding the ABC TLG. However, recent experiments have shown the
stability of TLG stacked with rhombohedral order, and the possibility of
opening a gap by applying a transverse electric field to the sample,\cite%
{LH11,BL11,JS11} what has activated the interest on TLG with this stacking
sequence. The Hamiltonian for ABC-stacked TLG around the $K$ point is%
\begin{equation}
H_{\mathbf{p}}=\left( 
\begin{array}{cccccc}
0 & v_{\mathrm{F}}p_{-} & 0 & 0 & 0 & 0 \\ 
v_{\mathrm{F}}p_{+} & 0 & \gamma _{1} & 0 & 0 & 0 \\ 
0 & \gamma _{1} & 0 & v_{\mathrm{F}}p_{-} & 0 & 0 \\ 
0 & 0 & v_{\mathrm{F}}p_{+} & 0 & \gamma _{1} & 0 \\ 
0 & 0 & 0 & \gamma _{1} & 0 & v_{\mathrm{F}}p_{-} \\ 
0 & 0 & 0 & 0 & v_{\mathrm{F}}p_{+} & 0%
\end{array}%
\right) .  \label{Eq:HamABC0}
\end{equation}%
\begin{figure*}[t]
\begin{center}
\mbox{
\includegraphics[width=8cm]{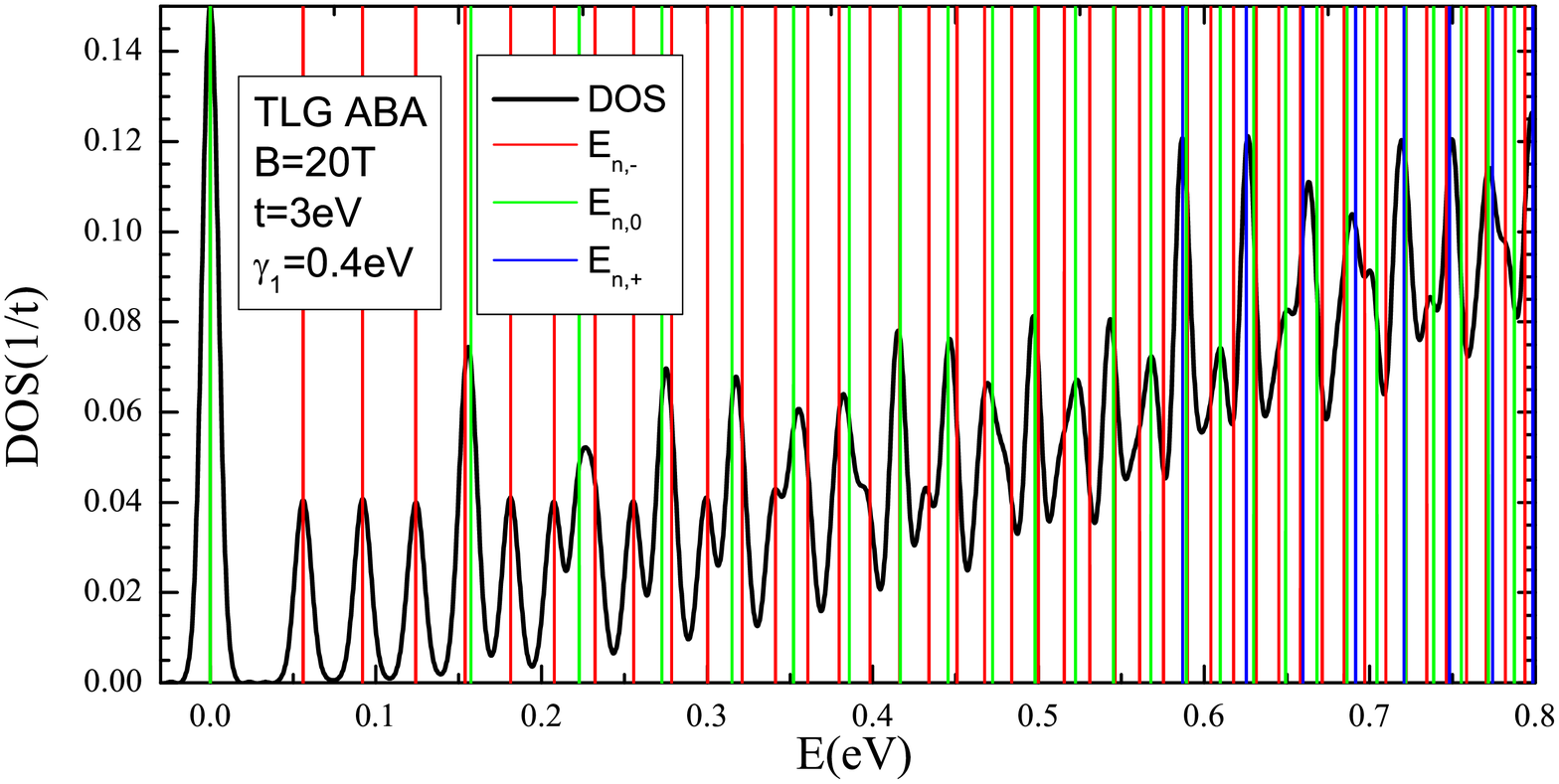}
\includegraphics[width=8cm]{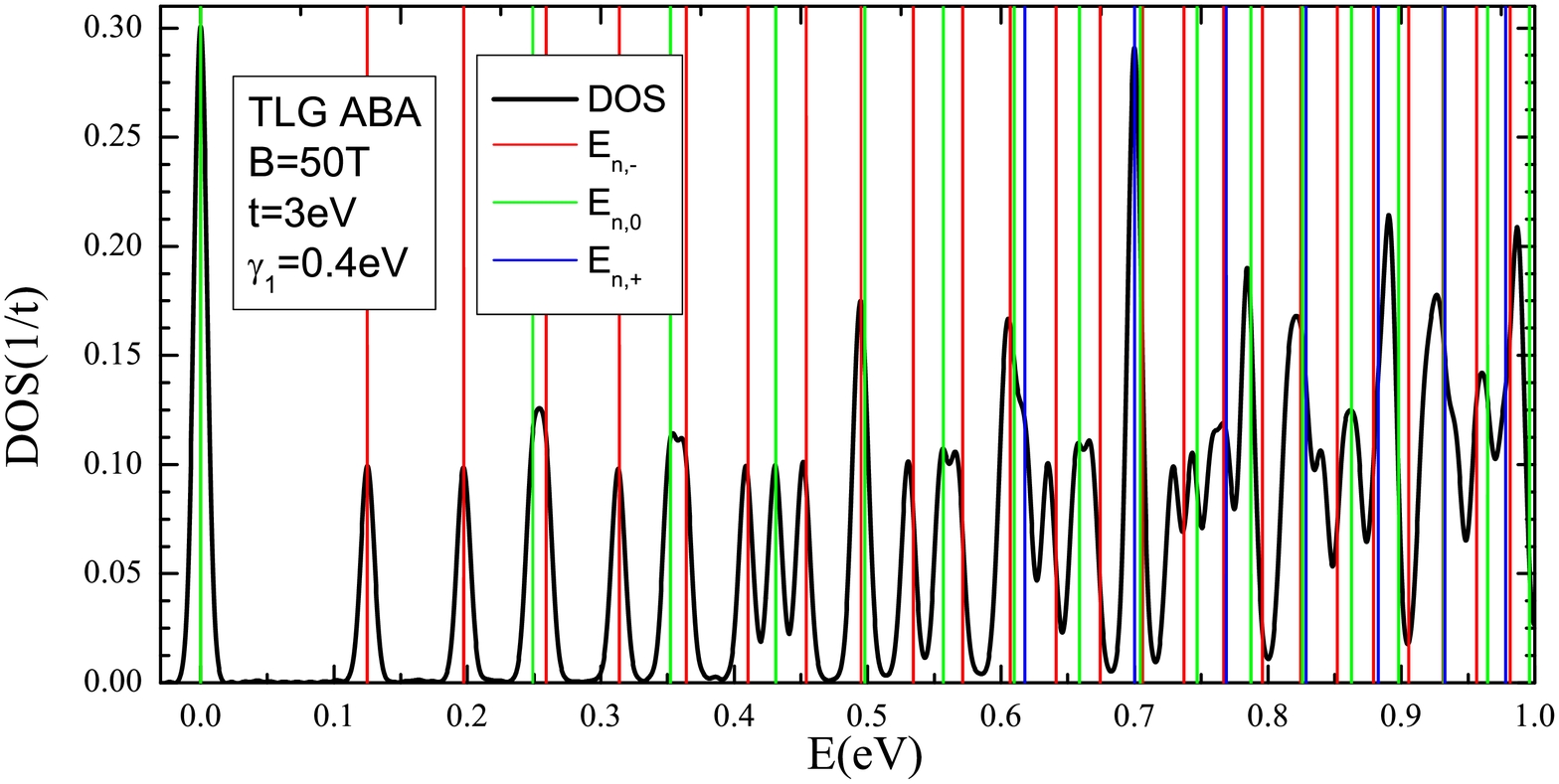}
} 
\mbox{
\includegraphics[width=8cm]{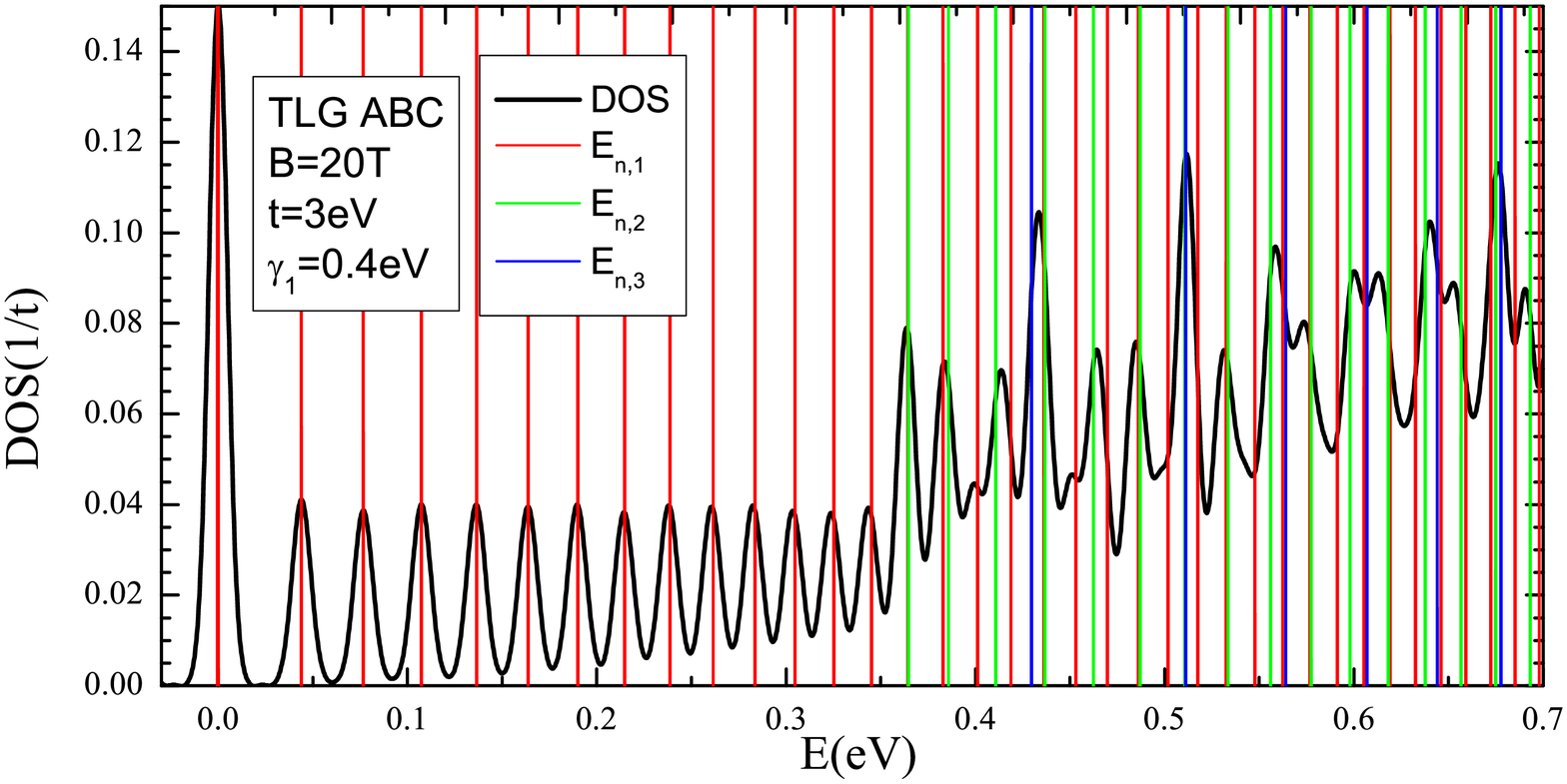}
\includegraphics[width=8cm]{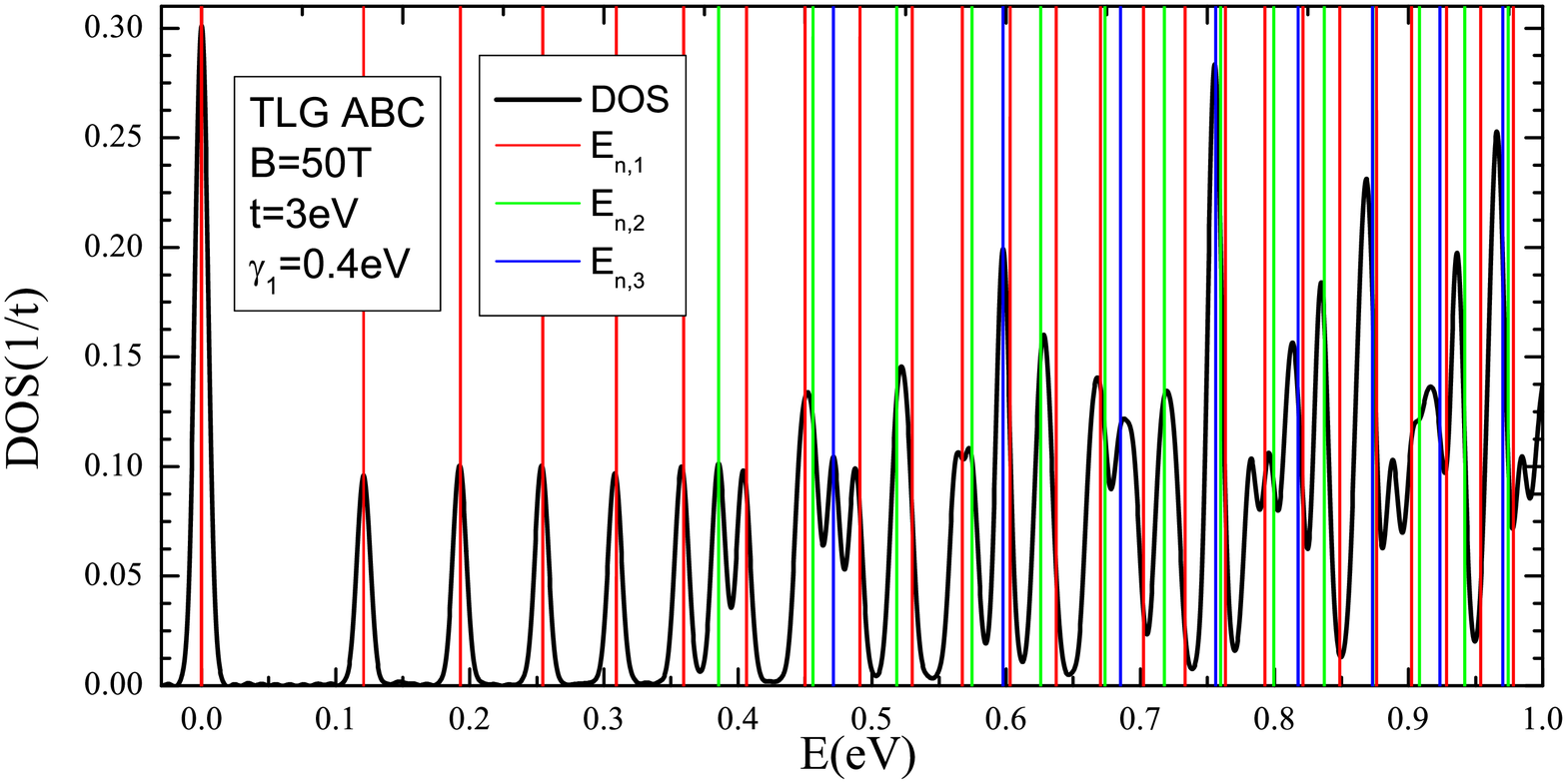}
}
\end{center}
\caption{(Color online) Comparation of the Landau level spectrum obtained
from the analytic expressions derived in the text (color lines) and the
numerical simulation (black lines) of ABA- and ABC-stacked trilayer
graphene. The sample used in the numerical simulations contains $3200\times
3200$ atomic sites in each layer, and we use the periodic boundary
conditions in the plane ($XY$) of graphene layers.}
\label{Fig:DOStrilayer}
\end{figure*}
The eigenvalues of Eq. (\ref{Eq:HamABC0}) leads, as shown in Fig. \ref%
{Fig:Bands} (b), to a low energy band structure that consists of a set of
six cubic bands, two of them touching each other at the $K$ point, and the
other four crossing at an energy $E=\pm \gamma _{1}$ above (below) the $K$
point. In the following we will obtain the LL spectrum for this case. In a
similar manner as for the ABA case, the six-components eigenstates of the
Hamiltonian for ABC-stacked TLG can be reconstructed as $\psi ={\Large [}%
c_{A_{1}}\varphi _{n-1,k}$, $c_{B_{1}}\varphi _{n,k}$, $c_{A_{2}}\varphi
_{n,k}$, $c_{B_{2}}\varphi _{n+1,k}$, $c_{A_{3}}\varphi _{n+1,k}$, $%
c_{B_{3}}\varphi _{n+2,k}{\Large ]}^{T},$ and the Hamiltonian matrix in this
case is ($n\geq 0$)%
\begin{equation}
\left( 
\begin{array}{cccccc}
0 & \Delta _{B}C_{1} & 0 & 0 & 0 & 0 \\ 
\Delta _{B}C_{1} & 0 & \gamma _{1} & 0 & 0 & 0 \\ 
0 & \gamma _{1} & 0 & \Delta _{B}C_{2} & 0 & 0 \\ 
0 & 0 & \Delta _{B}C_{2} & 0 & \gamma _{1} & 0 \\ 
0 & 0 & 0 & \gamma _{1} & 0 & \Delta _{B}C_{3} \\ 
0 & 0 & 0 & 0 & \Delta _{B}C_{3} & 0%
\end{array}%
\right) ,  \label{Eq:HamiltonianABC}
\end{equation}%
with $C_{1}=\sqrt{n}$, $C_{2}=\sqrt{n+1}$ and $C_{3}=\sqrt{n+2}$. The
eigenvalues of Eq. (\ref{Eq:HamiltonianABC}) are the solutions of the
equation 
\begin{equation}
E_{n}^{6}+bE_{n}^{4}+cE_{n}^{2}+d=0,  \label{eq_cubic}
\end{equation}%
where%
\begin{eqnarray}
b &=&-2\gamma _{1}^{2}-3\left( 1+n\right) \Delta _{B}^{2},  \notag
\label{Eq:bcd} \\
c &=&\gamma _{1}^{4}+2\left( 1+n\right) \gamma _{1}^{2}\Delta
_{B}^{2}+\left( 2+6n+3n^{2}\right) \Delta _{B}^{4}, \\
d &=&-n\left( n+1\right) \left( n+2\right) \Delta _{B}^{6},  \notag
\end{eqnarray}%
which leads to a LL spectrum for ABC-stacked TLG given by{\ \cite{BM96}}%
\begin{eqnarray}
E_{n,1} &=&\pm \sqrt{2\sqrt{Q}\cos \left( \frac{\theta +2\pi }{3}\right) -%
\frac{b}{3}},  \notag  \label{Eq:LLsABC} \\
E_{n,2} &=&\pm \sqrt{2\sqrt{Q}\cos \left( \frac{\theta +4\pi }{3}\right) -%
\frac{b}{3}}, \\
E_{n,3} &=&\pm \sqrt{2\sqrt{Q}\cos \left( \frac{\theta }{3}\right) -\frac{b}{%
3}},  \notag
\end{eqnarray}%
where%
\begin{eqnarray}
\theta &=&\cos ^{-1}\left( \frac{R}{\sqrt{Q^{3}}}\right) , \\
R &=&-\frac{b^{3}}{27}+\frac{bc}{6}-\frac{d}{2}, \\
Q &=&\frac{b^{2}}{9}-\frac{c}{3}.
\end{eqnarray}

In Eq.~(\ref{Eq:HamiltonianABC}), the Landau level index $n$ is required to
be nonnegative. However, notice that Eq. (\ref{Eq:HamiltonianABC}) admits
also eigenstates with real eigenvalues that contain components with $n=-1$.
The corresponding eigenenergies can be obtained by setting $C_{1}=-1$, $%
C_{2}=0$ and $C_{3}=1$ in Eq.~(\ref{Eq:HamiltonianABC}). This leads to three
twofold eigenvalues that complement Eq.~(\ref{Eq:LLsABC}) 
\begin{eqnarray}
E_{-1,1} &=&0,  \notag \\
E_{-1,3} &=&\pm \sqrt{\gamma _{1}^{2}+\Delta _{B}^{2}},  \notag
\end{eqnarray}%
where we label the contributions from the last two bands as $E_{-1,3}$,
because they have a similar field dependence as the $E_{n,3}$ LLs [see Fig. %
\ref{Fig:LLs}(d)].

In the low magnetic field limit, the Landau level spectrum for ABC-stacked
TLG can be approximated by\cite{GCP06,KM09} 
\begin{equation}
E_{n}\approx \pm \frac{\left( 2v_{F}^{2}/l_{B}^{2}\right) ^{3/2}}{\gamma
_{1}^{2}}\sqrt{n\left( n+1\right) \left( n+2\right) }.
\label{Eq:LLsABCapprox}
\end{equation}%
The positive energy part of the LL spectrum obtained from Eq. (\ref%
{Eq:LLsABC}) is represented in Fig. \ref{Fig:LLs}(c). One can distinguish
one set of LLs starting from zero energy, which correspond to the low energy
band that touches the Dirac point, plus two set of LLs at an energy $\sim
\gamma _{1}$ and which are related to the bands that cross at $\gamma _{1}$
[see Fig. \ref{Fig:Bands}(b)]. Whereas the low energy set of LLs can be
understood from a standard quantization of a low energy cubic band, the LLs
that appears at $E_{n}\sim \gamma _{1}$ deserve some discussion [see Fig. %
\ref{Fig:LLs}(d) for a zoom of the low field region of these states]. Most
saliently, the hybridization of the upper bands leads to two different sets
of LLs. One set of LLs [plotted in green color in Fig. \ref{Fig:LLs}%
(c)-(d)], associated to the inner branches of the hybridized bands [denoted
by the green lines in Fig. \ref{Fig:Bands}(b)], disperses with an energy $%
E_{n}>\gamma _{1}$ and it is quite similar to that of a SLG. The other set
of LLs, associated to the outer branches of the hybridized bands [denoted by
red lines in Fig. \ref{Fig:Bands}(b)], has an energy that first decrease
with $B$ until it reaches a minimum value, and then grows in energy as $B$
increases [see the lower set of LLs of Fig. \ref{Fig:LLs}(d), which are
colored in red]. This behavior is due to the cusp of this branch at $%
E=\gamma _{1}$, and resembles the saddle point of the bilayer graphene bands
in the presence of a transverse electric field. The effect of the
perpendicular electric field in BLG is to open a gap in the spectrum,
leading to Mexican hat like bands,\cite%
{McCann2006,CC07,Min2007,Zhang2009,Mak2009,Kuzmenko2009,Wang2010} with the
corresponding anomalous LL quantization of the band.\cite{PPV07,ZFA10,K11}
Therefore, the LLs associated to the quantization of the lower branches of
the hybridized bands in ABC TLG can be obtained, in a first approximation,
by using the semiclassical approximation used in Ref. \onlinecite{ZFA10} for
a biased bilayer graphene. The degeneracy of zero-order Landau level in ABC
TLG is three times larger than SLG. This result remains correct also for the
case of inhomogenous magnetic field as follows from the index theorem. \cite%
{Katsnelson2008}

\section{Density of states from a full $\protect\pi $-band tight-binding
model}

\label{Sec:Num}

\begin{figure}[t]
\begin{center}
\includegraphics[width=7cm]{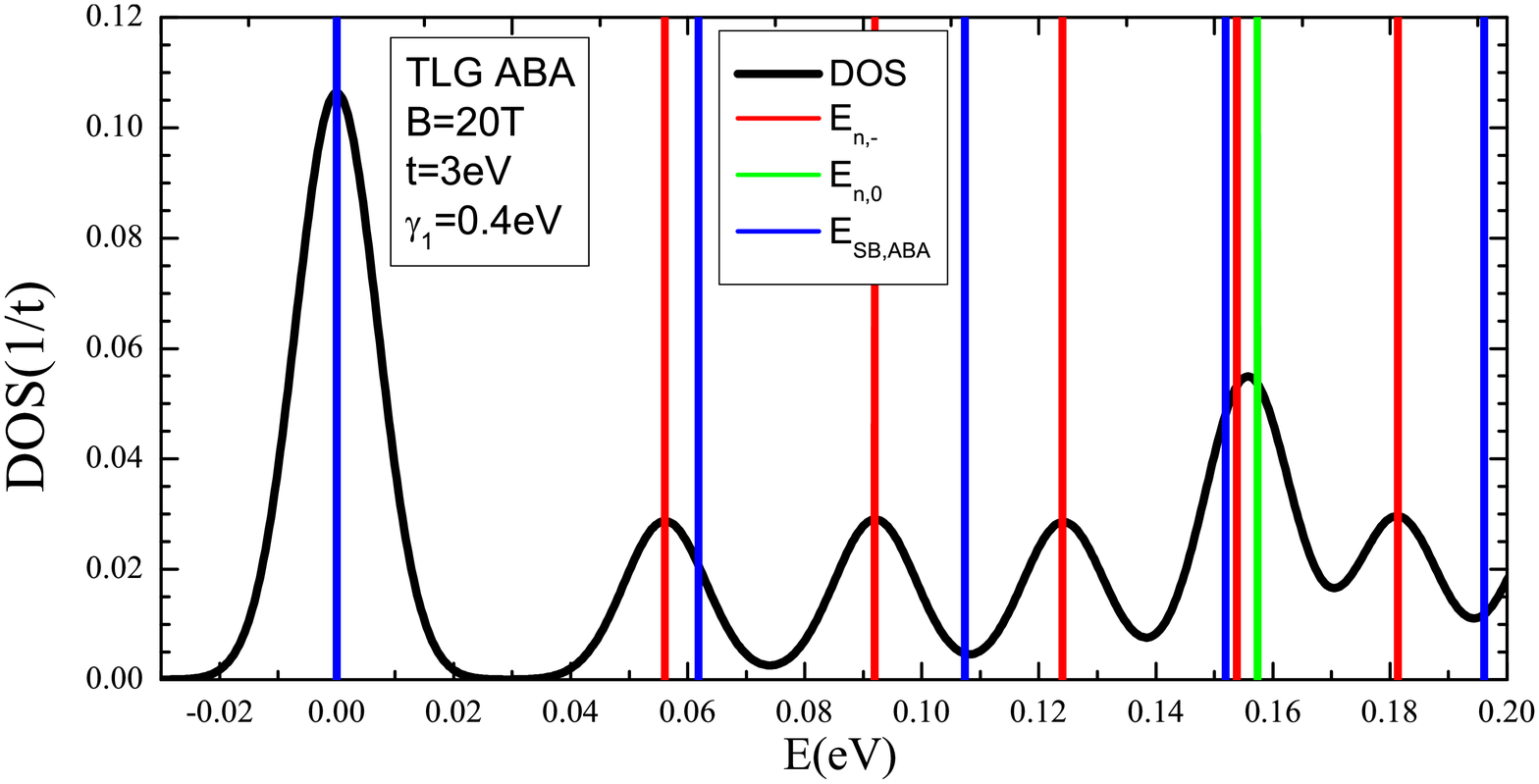} %
\includegraphics[width=7cm]{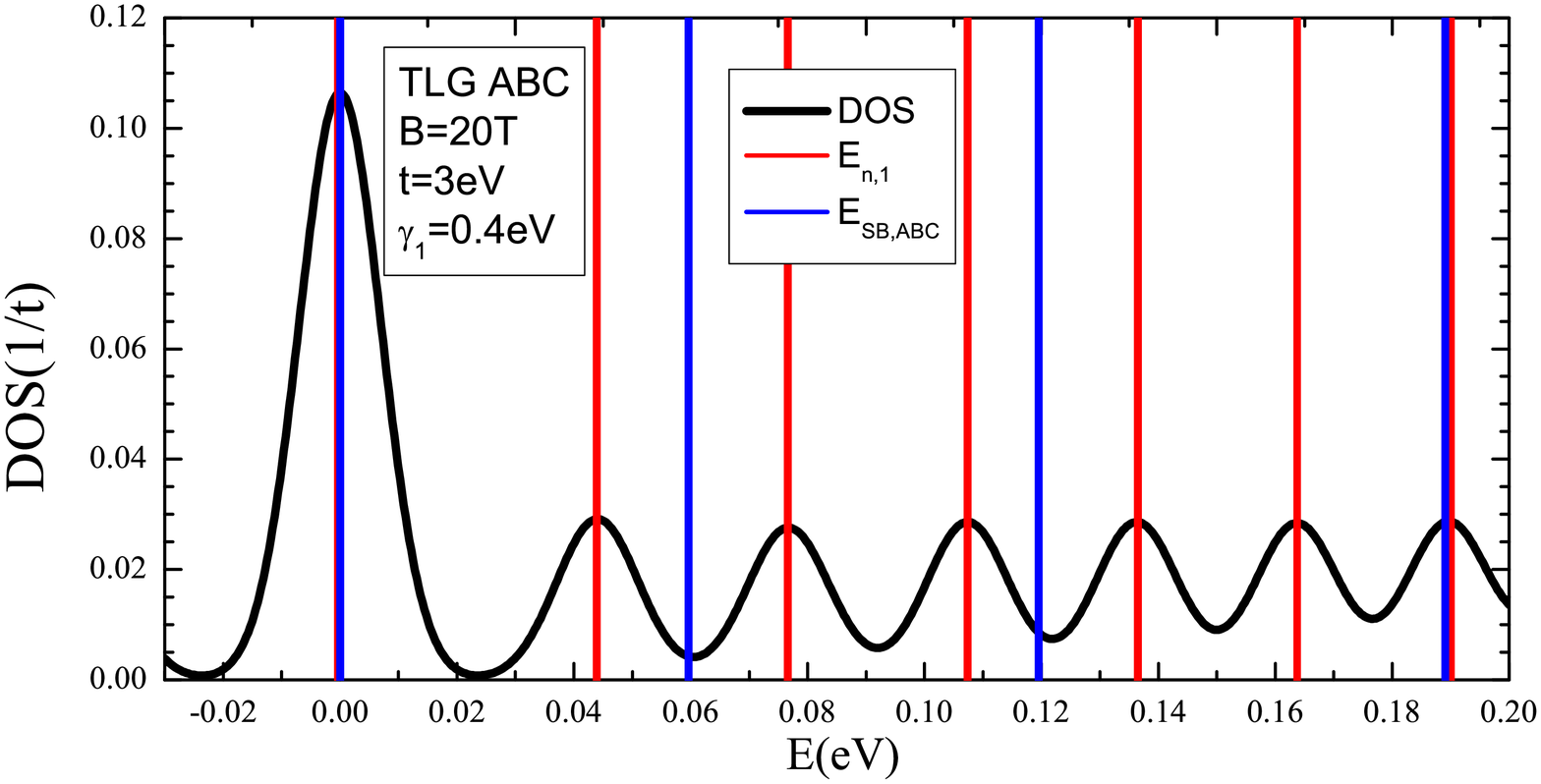} 
\end{center}
\caption{ (Color online) Landau level spectrum of (a) ABA- and (b)
ABC-stacked trilayer graphene at $B=20$~T obtained from the numerical
solution of the TDSE using a $\protect\pi$-band tight-binding model (black
lines). In (a) the TDSE results is compared to the results from the analytic
expression $E_{n,-}$ (red vertical lines) and $E_{n,0}$ (green vertical
lines) from Eq. (\protect\ref{Eq:LLsABA})-(\protect\ref{Eq:DispABA-SLG}),
and to the approximation Eq. (\protect\ref{Eq:LLsABAapprox}). In (b), the
TDSE DOS are compared to the analytic result for $E_{n,1}$ from Eq. (\protect
\ref{Eq:LLsABC}), and to the approximation Eq. (\protect\ref{Eq:LLsABCapprox}%
) (blue line).}
\label{Fig:DOSapprox}
\end{figure}

In order to check the range of validity of the analytic expressions obtained
in Sec. \ref{Sec:LLspectrum}, in this section we compare the LLs obtained
from the equations (\ref{Eq:LLsABA})-(\ref{Eq:DispABA-SLG}) and (\ref%
{Eq:LLsABC}) for the low energy spectrum of ABA- and ABC-stacked TLG,
respectively, to the density of states (DOS) obtained numerically by solving
the time-dependent Schr\"{o}dinger equation (TDSE) on a honeycomb lattice in
the framework of a $\pi $-band tight-binding model.\cite{HR00,YRK10,YRK10b}
The effect of an external magnetic field is considered by means of a Peierls
substitution 
\begin{equation}
t_{mn}\rightarrow t_{mn}e^{ie\int_{m}^{n}\mathbf{A}\cdot d\mathbf{l}},
\end{equation}%
where $t_{mn}$ is the hopping amplitude between sites $m$ and $n$ of the
honeycomb lattice, and $\int_{m}^{n}\mathbf{A}\cdot d\mathbf{l}$ is the line
integral of the vector potential. A numerical study of the
magneto-electronic properties of ABC TLG has been also reported in Ref. %
\onlinecite{Ho2011}. In Fig. \ref{Fig:DOStrilayer} we compare our analytic
results of Eqs. (\ref{Eq:LLsABA})-(\ref{Eq:DispABA-SLG}) and (\ref{Eq:LLsABC}%
) with the numerical TDSE results for the DOS, for two different values of
magnetic field. We find a very good agreement between analytic and
tight-binding results up to an energy of $\sim 0.5$~eV. Notice that when a
LL crossing occurs, for example of a SLG-like LL crossing with a BLG-like LL
in ABA TLG, this leads to an increase of the peak in the DOS. This is, e.
g., the reason for the enhanced peaks at $E\approx 0.5$~eV and $E\approx 0.7$%
~eV in Fig. \ref{Fig:DOStrilayer}(b), as it can be deduced by following the
LL spectrum of Fig. \ref{Fig:LLs}(a) at $B=50$~T. Far from the neutral
point, at an energy $E\gtrsim 0.5$~eV the analytic results are shifted to
the right of the spectrum, as compared with the numerical TDSE results (see
e. g. the peaks corresponding to $E_{n,-}$ for ABA- and $E_{n,1} $ for
ABC-stacked TLG, represented by the red vertical lines in Fig. \ref%
{Fig:DOStrilayer}). This is due to the fact that the dispersion relation for
SLG is not linear anymore, so that higher order terms should be included for
a precise reproduction of the position of the LLs.

\begin{figure}[t]
\begin{center}
\includegraphics[width=8cm]{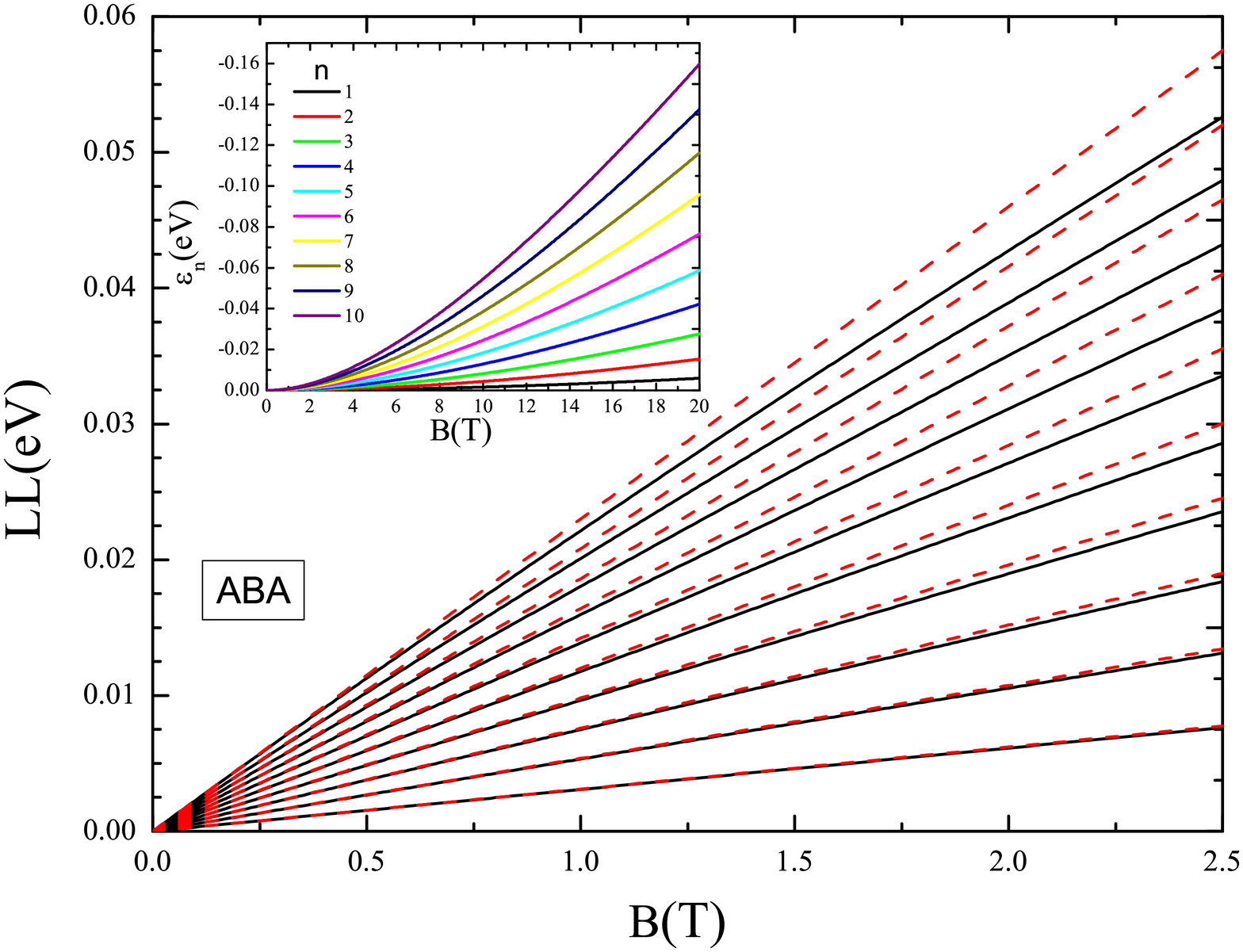} %
\includegraphics[width=8cm]{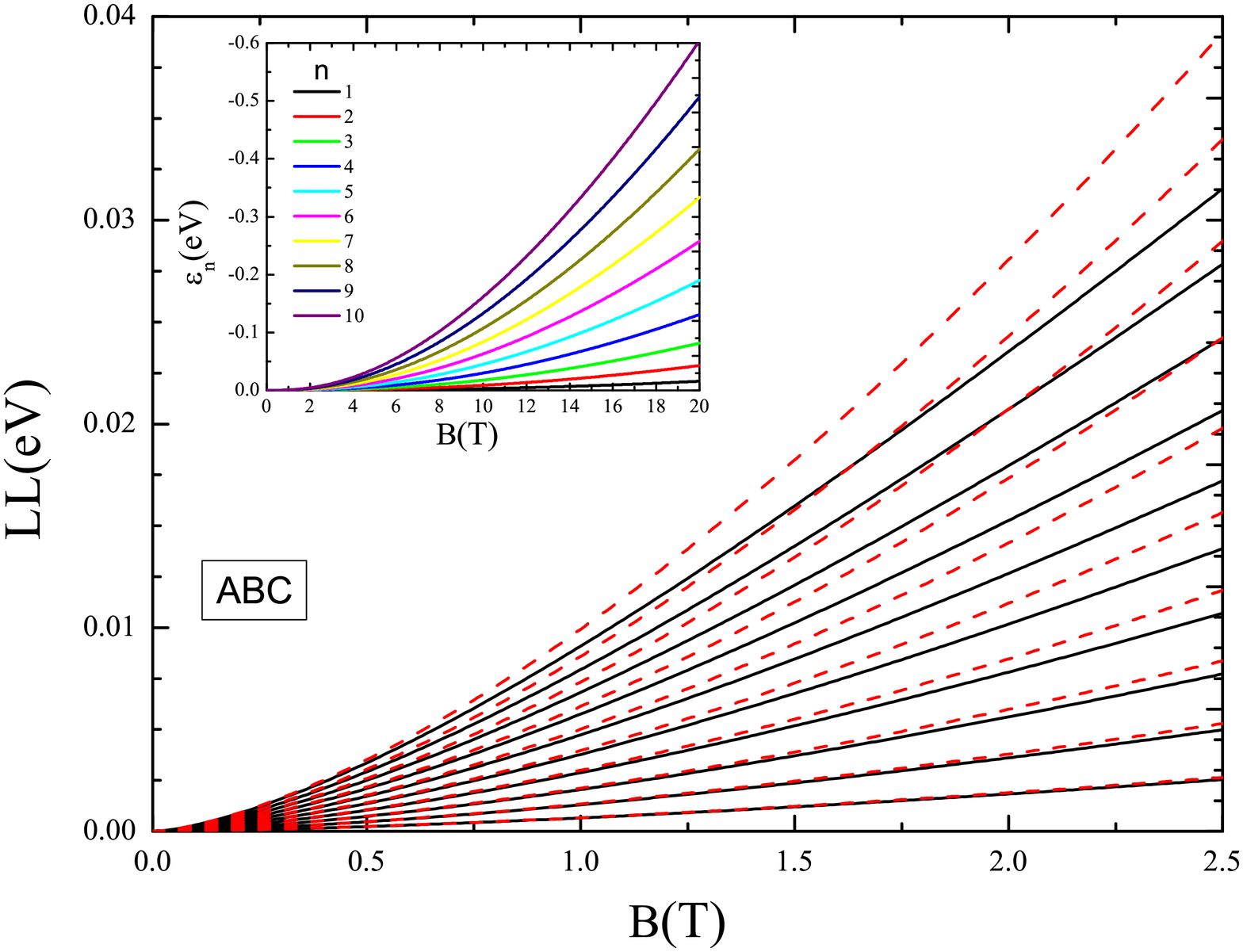} 
\end{center}
\caption{ (Color online) Comparation of the analytic results for the first
ten Landau levels obtained from Eqs. (\protect\ref{Eq:LLsABA})-(\protect\ref%
{Eq:DispABA-SLG}) and (\protect\ref{Eq:LLsABC}) for ABA- and ABC-stacked
graphene, respectively (black solid lines) and the approximations Eqs. (%
\protect\ref{Eq:LLsABAapprox}) and (\protect\ref{Eq:LLsABCapprox}) (red
dashed lines). The inset panels are the difference between Eq. (\protect\ref%
{Eq:LLsABA})-(\protect\ref{Eq:LLsABC}) and the commonly used approximations
Eqs. (\protect\ref{Eq:LLsABAapprox})-(\protect\ref{Eq:LLsABCapprox}). Notice
the different range of magnetic fields used in the inset with respect to the
main figures.}
\label{Fig:DiffLLs}
\end{figure}

It is interesting also to check the range of validity of the most commonly
used approximated expressions for the LL spectrum of TLG [Eq. (\ref%
{Eq:LLsABAapprox}) for ABA and Eq. (\ref{Eq:LLsABCapprox}) for ABC].
Contrary to single layer graphene, for which the LL spectrum behaves as $%
\sqrt{Bn}$ up to rather high energies (in Ref. \onlinecite{PH08} it was
reported a deviation of only $\sim 40$~meV at an energy of $1.25$~eV), the $B%
\sqrt{n(n+1)}$ behavior of the BLG-like LLs of ABA TLG as well as the $%
B^{3/2}\sqrt{n(n+1)(n+2)}$ behavior of ABC TLG are valid only in a rather
reduced range of energies in the spectrum. In fact, we see in Fig. \ref%
{Fig:DOSapprox} that, for the moderate value of magnetic field used for this
plot ($B=20$~T) the approximations Eqs. (\ref{Eq:LLsABAapprox}) and (\ref%
{Eq:LLsABCapprox}) fail to capture accurately even the second LL of the
spectrum. The deviation is especially important for ABC trilayer graphene,
as seen in Fig. \ref{Fig:DiffLLs}, where one can see that there are
deviations of hundreds of meV between the two results already for low LLs at
some intermediate values of magnetic field $\sim 15-20$~T. This is somehow
expected since recent cyclotron resonance experiments\cite{HS08,OP11} on
bilayer graphene required the use of the equivalent expression for BLG of
Eq. (\ref{Eq:DispABA-BLG}), that we have obtained for the BLG-like bands of
ABA TLG. Indeed, a good fitting (apart from some possible many-body
corrections\cite{RFG10,S11}) of the magneto-optical experiments on BLG was
achieved by using an expression similar to Eq. (\ref{Eq:DispABA-BLG}), with
the only tight-binding parameters $\gamma_0\equiv t$ and $\gamma_1$.
Therefore, we expect that the analytic expressions Eqs. (\ref{Eq:LLsABA})-(%
\ref{Eq:DispABA-SLG}) and (\ref{Eq:LLsABC}) that we have obtained can be
useful when analyzing future cyclotron resonance experiments of ABA- and
ABC-stacked trilayer graphene.

\begin{figure}[t]
\begin{center}
\includegraphics[width=8cm]{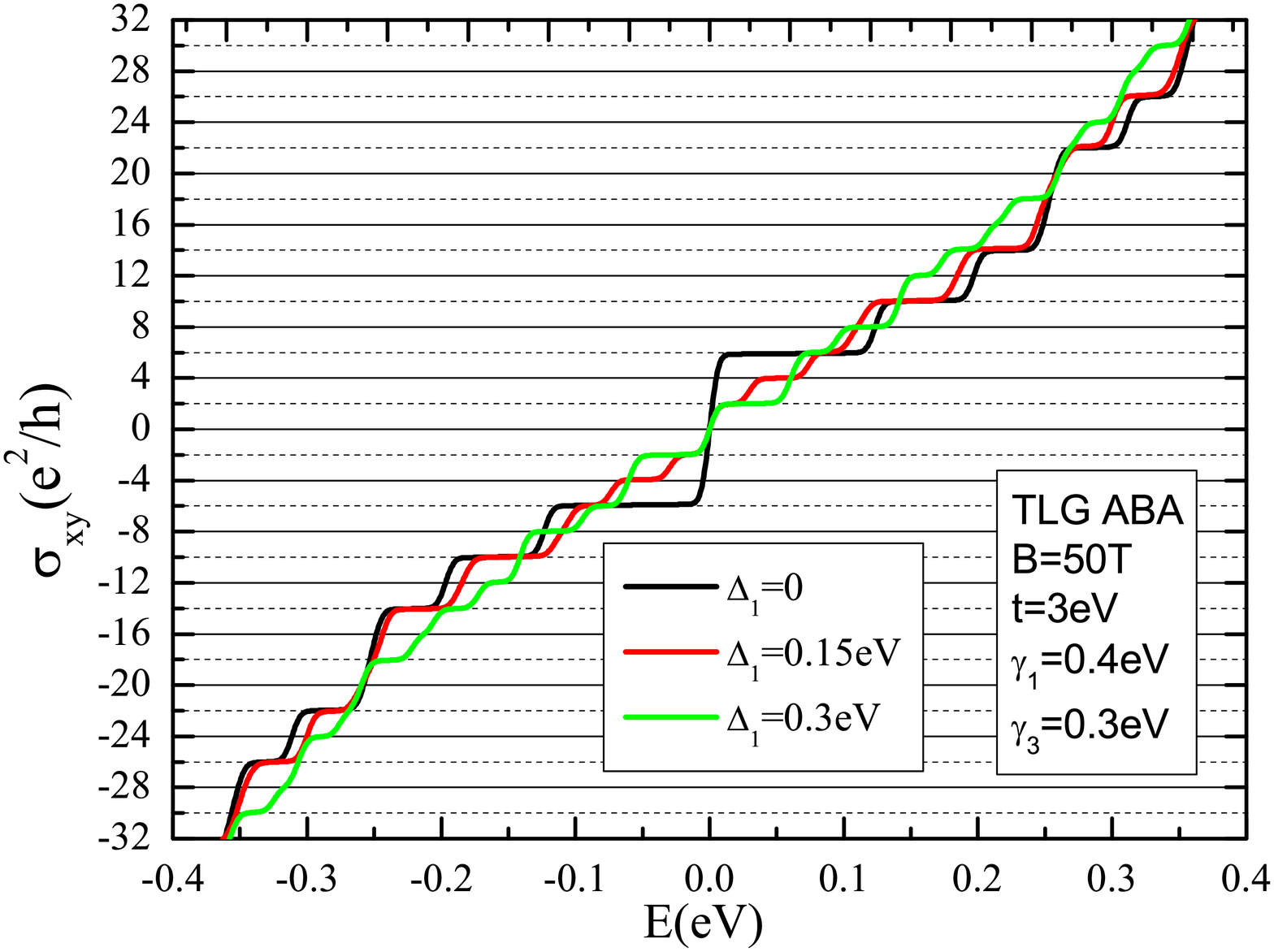} 
\includegraphics[width=8cm]{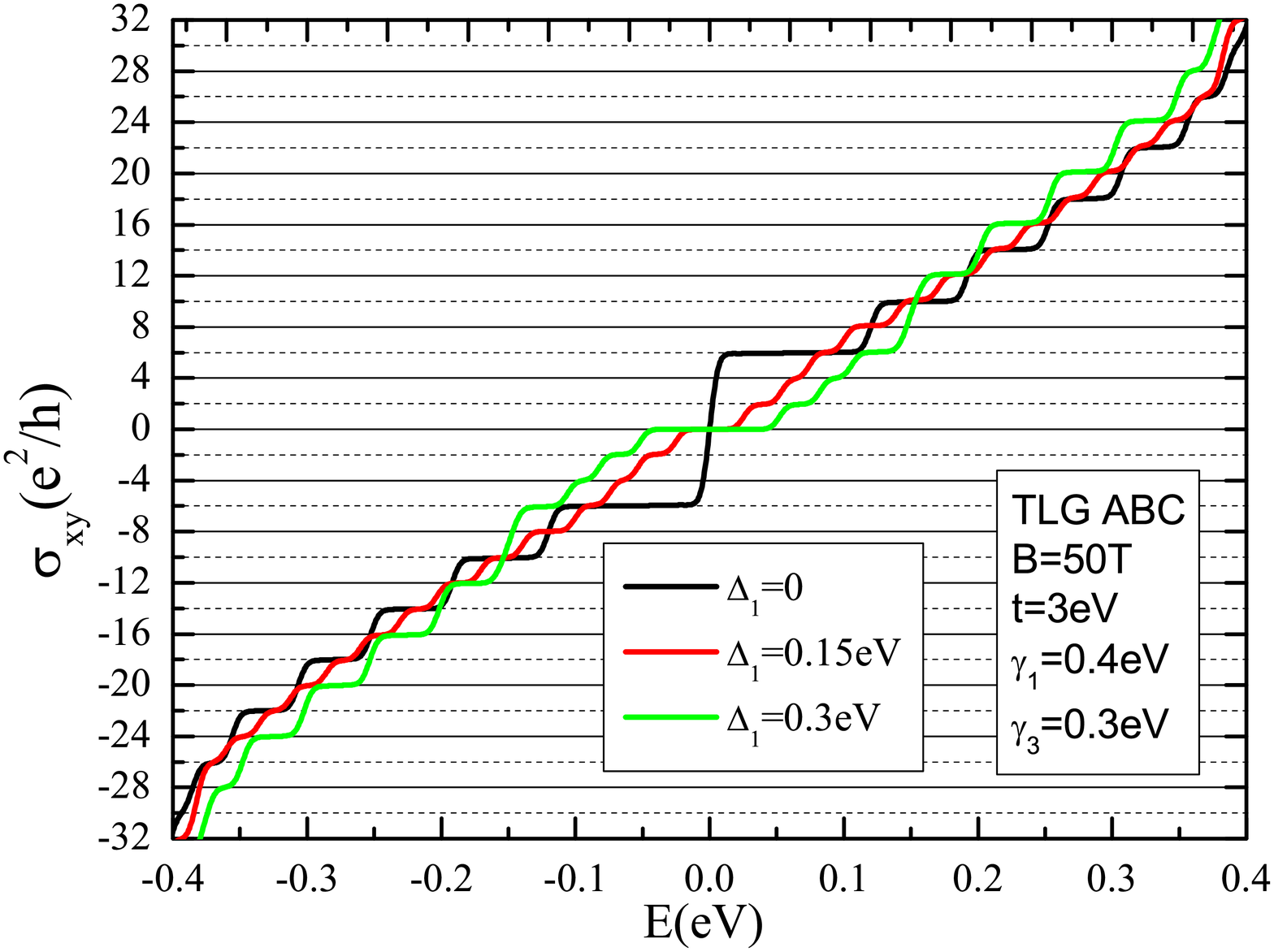} 
\end{center}
\caption{ (Color online) Hall conductivity of ABA- and ABC-stacked trilayer
graphene with different values of $\Delta_1$ induced by a transverse
electric field.}
\label{Fig:sigma}
\end{figure}

Furthermore, motivated by recent transport measurements on TLG, which have
revealed the strongly stacking dependent quantum Hall effect in this
material,\cite{TJ11,BL11,LH11,KR11,ZZ11,JS11} we have calculated the Hall
conductivity for the two stacking sequences of TLG, considering also the
effect of a transverse electric field in the spectrum. Here the Hall
conductivity $\sigma_{xy}$ is calculated by using the Kubo formula \cite{K57}%
\begin{equation}
\sigma _{xy}=-\frac{n_sec}{B}+\Delta \sigma _{xy},
\end{equation}%
where the charge density $n_s=\int_{0}^{E}\rho \left( E\right) dE$ is
obtained by integration of the DOS $\rho(E)$ calculated from the TDSE and $%
\pi$-band tight-binding method, and $\Delta \sigma _{xy}$ is a correction
due to scattering of electrons with impurities,\cite{YRK10} and which is
zero in the clean limit considered here. In Fig. \ref{Fig:sigma}, we show
the Hall conductivity of ABA- and ABC-stacked TLG with or without an
external electric field. In the absence of any bias, the Hall conductivity
for the two cases is similar, with plateaus at $\nu=\pm 6,\pm 10,\pm 14,...$%
. However, the structure of $\sigma_{xy}$ is different when we consider the
effect of a transverse electric field, which is accounted for here by adding
a different (nonzero) on-site potential on the top and the bottom layers,
namely, $\Delta _{1}/2$ on the top layer and $-\Delta _{1}/2$ on the bottom
layer. The main difference between ABA- and ABC-stacked TLG in the presence
of a transverse bias is that it leads to a gap opening in the case of
ABC-stacking, while the ABA-stacked TLG remains gapless, as it has been
observed experimentally.\cite{BL11} In fact, the opening of the gap and the
corresponding insulating state leads to the appearance of a zero energy
plateau in the Hall conductivity in ABC TLG, plateau which is absent in ABA
TLG, as shown in Fig. \ref{Fig:sigma} for different values of $\Delta_1$. On
the other hand, the position of the plateaus depends very much on the value
of the induced difference potential $\Delta_1$. For a small bias leading to $%
\Delta_1=0.15$~eV, we find plateaus for ABA TLG at $\nu=\pm 2,\pm 4,\pm
6,\pm10,\pm14,...$, whereas a higher value, $\Delta_1=0.3$~eV leads to
plateaus at $\nu=\pm2,\pm6,\pm8,\pm12,\pm14,...$. On the other hand, whereas 
$\Delta_1=0.15$~eV leads to plateaus for ABC at all even values of $\nu$
(including $\nu=0$), some of the plateaus are missing for a higher value of
bias, $\Delta_1=0.3$~eV, for which we find plateaus at $\nu=0,\pm 2,\pm
4,\pm 6,\pm 12,\pm 16,...$. 
In fact, a more deep understanding of the Hall conductivity of TLG would
require further analysis, which is beyond the scope of this work.
Furthermore, we emphasize that even experimentally, there is no consensus so
far about the structure of the quantum Hall plateaus in trilayer graphene,
having been found different structures for almost every transport
measurement.\cite{BL11,KR11,ZZ11,JS11}

\section{Conclusions}

\label{Sec:Concl}

In conclusion, we have derived analytic expressions for the Landau level
spectrum of trilayer graphene. The two stable stacking sequence, ABA
(Bernal) and ABC (rhombohedral) have been considered. The LL spectrum for
ABA TLG is composed by a set of bilayer graphene-like LLs, which disperse at
low energies as $B$, and a set of single layer graphene-like LLs, which
disperse as $\sqrt{B}$. The different character of the bands lead to a
series of LL crossings, which has been observed experimentally.\cite{TJ11}
On the other hand, the six cubic bands of ABC TLG leads to a rather peculiar
LL quantization of the spectrum. Whereas the bands that touch the Dirac
point lead to a set of $B^{3/2}$ LLs, the hybridization between the two
bands that cross each other at $E=\gamma _{1}$ leads to one set of massless
like LLs (with energy $E\geq \gamma _{1}$), and a set of LLs which present a
minimum and then grows with $B$, associated to the lower branch of the
hybridized bands. The presence of the minimum on this set of LLs is
associated to the presence of a cusp in this branch of the spectrum, in a
similar manner as the Mexican hat like dispersion of a biased bilayer
graphene.

The range of validity of our analytical results is checked by comparing the
LL spectrum obtained in the continuum approximation to the density of states
obtained from the numerical solution of the time dependent Schr\"{o}dinger
equation of a $\pi$-band tight-binding model on the honeycomb lattice. We
find very good agreement between the numerical solution and the analytic
approximation for the spectrum up to an energy of $\sim 500$~meV. However,
we show that the most commonly used approximations for the spectrum of TLG,
for which the BLG-like LLs of ABA TLG disperse as $B\sqrt{n(n+1)}$ and the
LLs for TLG disperse as $B^{3/2}\sqrt{n(n+1)(n+2)}$, fail to capture even
the lower LLs already for moderate magnetic fields of $\sim 20$~T.
Therefore, we believe that our results maybe useful for the analysis of
future magneto-optical measurements, which has been successfully applied to
study the LL spectrum of SLG\cite{SH06,JS07} and BLG.\cite{HS08,OP11}

Finally, we have calculated the Hall conductivity of TLG by means of the
Kubo formula. The inclusion of a transverse electric field leads to a gap
opening in ABC TLG, whereas ABA TLG remains metallic. This effect is seen by
the appearance of a zero energy plateau only for ABC stacking, in agreement
with recent transport experiments.\cite{BL11,KR11,ZZ11,JS11}

\section{Acknowledgement}

The authors thank useful conversations with E. V. Castro, E. Cappelluti and
F. Guinea. The support by the Stichting Fundamenteel Onderzoek der Materie
(FOM) and the Netherlands National Computing Facilities foundation (NCF) are
acknowledged. We thank the EU-India FP-7 collaboration under MONAMI and the
grant CONSOLIDER CSD2007-00010.

\appendix

\section{Band structure of ABA and ABC trilayer graphene in the absence of
magnetic field}

In the absence of a magnetic field, the Hamiltonian of ABA-stacked TLG
around the $K$ point is given in Eq. (\ref{Eq:HamABA0}), {with eigenenergies
given by}%
\begin{eqnarray}
E_{s} &=&\pm {\Large [}\gamma _{1}^{2}+v_{F}^{2}k^{2}+s\sqrt{\gamma
_{1}^{4}+2\gamma _{1}^{2}v_{F}^{2}k^{2}}{\Large ]}^{1/2},\text{ }s=\pm 1 
\notag \\
E_{0} &=&\pm v_{F}k.
\end{eqnarray}

{Similarly, for ABC-stacked TLG, the Hamiltonian Eq. (\ref{Eq:HamABC0})
leads to the eigenvalue problem} 
\begin{eqnarray}
&&E^{6}-\left( 2\gamma _{1}^{2}+3v_{F}^{2}k^{2}\right) E^{4}+  \notag \\
&&\left( \gamma _{1}^{4}+2\gamma
_{1}^{2}v_{F}^{2}k^{2}+3v_{F}^{4}k^{4}\right) E^{2}-v_{F}^{6}k^{6}=0,
\label{bandabc_b0}
\end{eqnarray}%
the solutions of which take the form of Eq.~(\ref{Eq:LLsABC}) with the new
quantities $b=$ $-2\gamma _{1}^{2}-3v_{F}^{2}k^{2}$, $c=\gamma
_{1}^{4}+2\gamma _{1}^{2}v_{F}^{2}k^{2}+3v_{F}^{4}k^{4}$ and $d=$\ $%
-v_{F}^{6}k^{6}$. In fact, Eq.~(\ref{bandabc_b0}) can be decomposed into the
two equations:%
\begin{equation}
E^{3}+v_{F}kE^{2}-\left( \gamma _{1}^{2}+v_{F}^{2}k^{2}\right)
E-v_{F}^{3}k^{3}=0,  \label{bandabc_b0a}
\end{equation}%
\begin{equation}
E^{3}-v_{F}kE^{2}-\left( \gamma _{1}^{2}+v_{F}^{2}k^{2}\right)
E+v_{F}^{3}k^{3}=0,  \label{bandabc_b0b}
\end{equation}%
the solutions of which are%
\begin{eqnarray}
E_{\alpha ,s} &=&2\sqrt{Q}\cos \left( \frac{\theta +2\pi }{3}\right) -s\frac{%
v_{F}k}{3},  \notag \\
E_{\beta ,s} &=&2\sqrt{Q}\cos \left( \frac{\theta +4\pi }{3}\right) -s\frac{%
v_{F}k}{3}, \\
E_{\gamma ,s} &=&2\sqrt{Q}\cos \left( \frac{\theta }{3}\right) -s\frac{v_{F}k%
}{3},  \notag
\end{eqnarray}%
where $s=\pm 1$ correspond to the solutions of Eq.~(\ref{bandabc_b0a}) and (%
\ref{bandabc_b0b}) respectively, in terms of the new parameters%
\begin{eqnarray}
\theta  &=&\cos ^{-1}\left( \frac{sR}{\sqrt{Q^{3}}}\right) , \\
R &=&\frac{8v_{F}^{3}k^{3}}{27}-\frac{v_{F}k\gamma _{1}^{2}}{6}, \\
Q &=&\frac{3\gamma _{1}^{2}+4v_{F}^{2}k^{2}}{9}.
\end{eqnarray}

\section{Wave functions of ABA trilayer graphene}

From the matrix Hamiltonian Eq. (\ref{Eq:Ha}) one can calculate the
eigenstates of the ABA TLG. They are given by

\begin{widetext}
\begin{eqnarray}
\psi _{n,s}\left( x,y\right)  &=&\left[ 
\begin{array}{c}
\pm \left\{ \frac{n\text{$\Delta _{B}$}^{2}-E_{n,s}^{2}}{\sqrt{n}E_{n,s}%
\text{$\Delta _{B}$}}-\frac{E_{n,s}}{\sqrt{n}\text{$\Delta _{B}$}}\left[ 1-%
\frac{(1+n)\text{$\Delta _{B}^{2}$}\left( n\text{ $\Delta _{B}$}%
^{2}-E_{n,s}^{2}\right) }{\text{$\gamma _{1}$}^{2}E_{n,s}^{2}}\pm \frac{%
E_{n,s}^{2}-n\text{$\Delta _{B}$}^{2}}{\text{$\gamma _{1}$}^{2}}\right]
\right\} \varphi _{n-1,k}(x,y) \\ 
\left[ -1+\frac{(1+n)\text{$\Delta _{B}^{2}$}\left( n\text{ $\Delta _{B}$}%
^{2}-E_{n,s}^{2}\right) }{\text{$\gamma _{1}$}^{2}E_{n,s}^{2}}\pm \frac{n%
\text{$\Delta _{B}$}^{2}-E_{n,s}^{2}}{\text{$\gamma _{1}$}^{2}}\right]
\varphi _{n,k}(x,y) \\ 
\pm \left( \frac{E_{n,s}}{\text{$\gamma _{1}$}}-\frac{n\text{$\Delta _{B}$}%
^{2}}{\text{$\gamma _{1}$}E_{n,s}}\right) \varphi _{n,k}(x,y) \\ 
\left( \frac{\sqrt{1+n}\text{$\Delta _{B}$}}{\text{$\gamma _{1}$}}-\frac{n%
\sqrt{1+n}\text{$\Delta _{B}$}^{3}}{\text{$\gamma _{1}$}E_{n,s}^{2}}\right)
\varphi _{n+1,k}(x,y) \\ 
\pm \frac{\sqrt{n}\text{$\Delta _{B}$}}{E_{n,s}}\varphi _{n+1,k}(x,y) \\ 
\varphi _{n+2,k}(x,y)%
\end{array}%
\right] ,  \notag \\
&&
\end{eqnarray}%
and%
\begin{equation}
\psi _{n,0}=\left[ 
\begin{array}{c}
\mp \varphi _{n-1,k}(x,y) \\ 
-\varphi _{n,k}(x,y) \\ 
0 \\ 
0 \\ 
\pm \varphi _{n+1,k}(x,y) \\ 
\varphi _{n+2,k}(x,y)%
\end{array}%
\right] .
\end{equation}%
\end{widetext}

Notice that the states with the eigenvalues $E_{n,0}$ are the surface states
which are located only on the top and bottom layers, and these surface
states in each layer have the same expressions as the single-layer graphene{.%
}

\section{Effect of $\protect\gamma_3$ in the DOS}

\label{App:gamma3}

\begin{figure*}[t]
\begin{center}
\mbox{
\includegraphics[width=9cm]{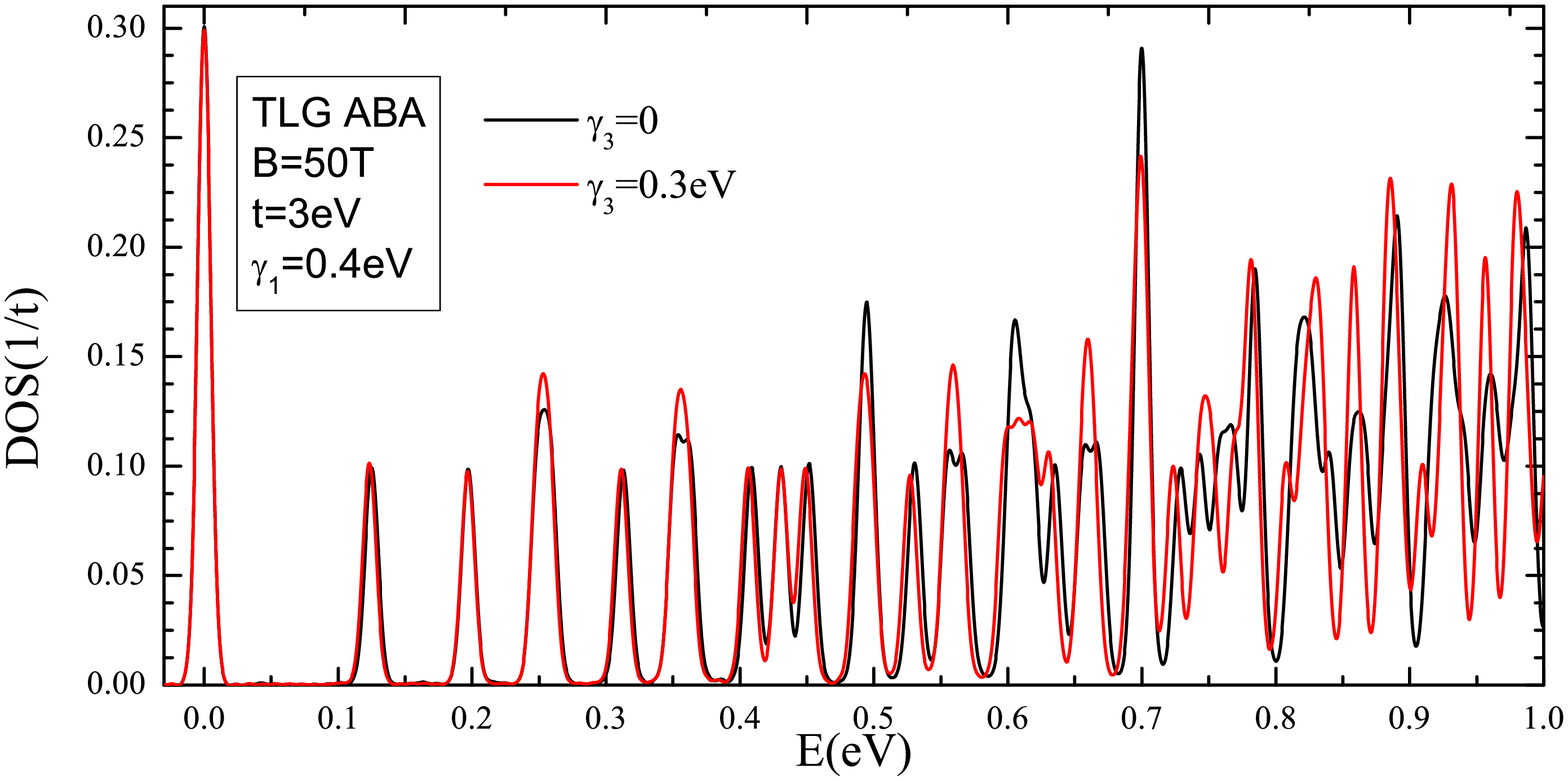}
\includegraphics[width=6cm]{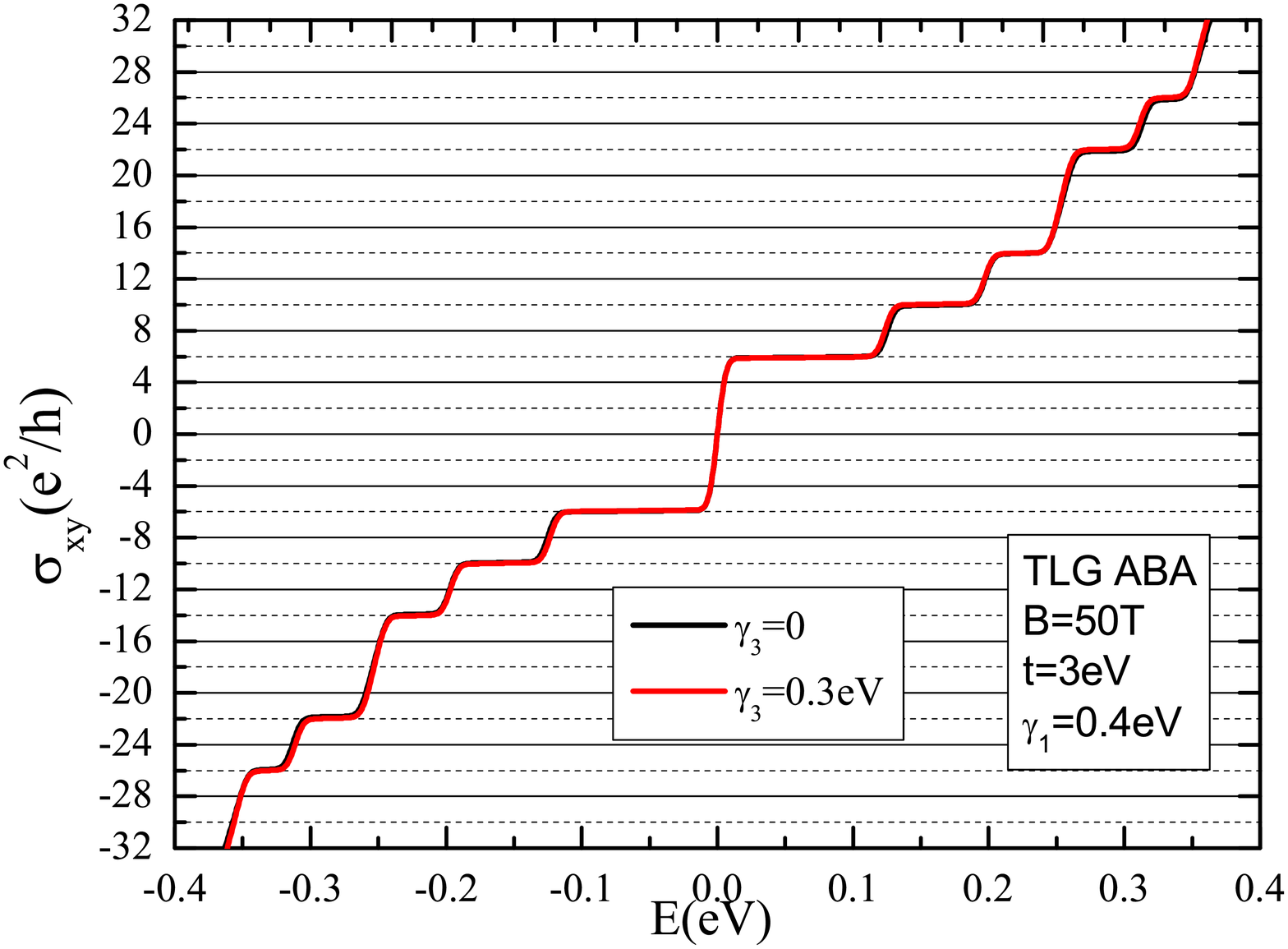}
} 
\mbox{
\includegraphics[width=9cm]{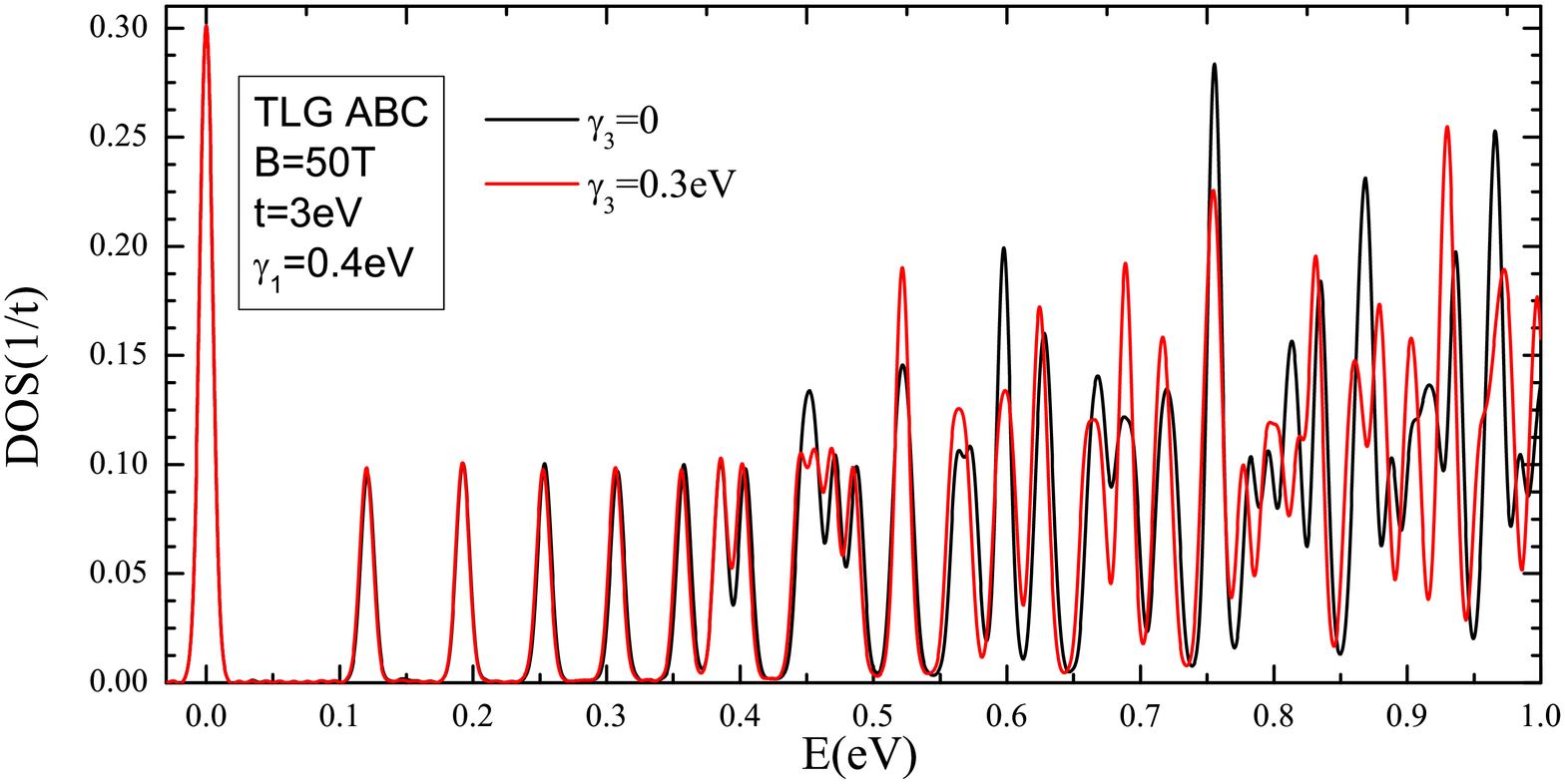}
\includegraphics[width=6cm]{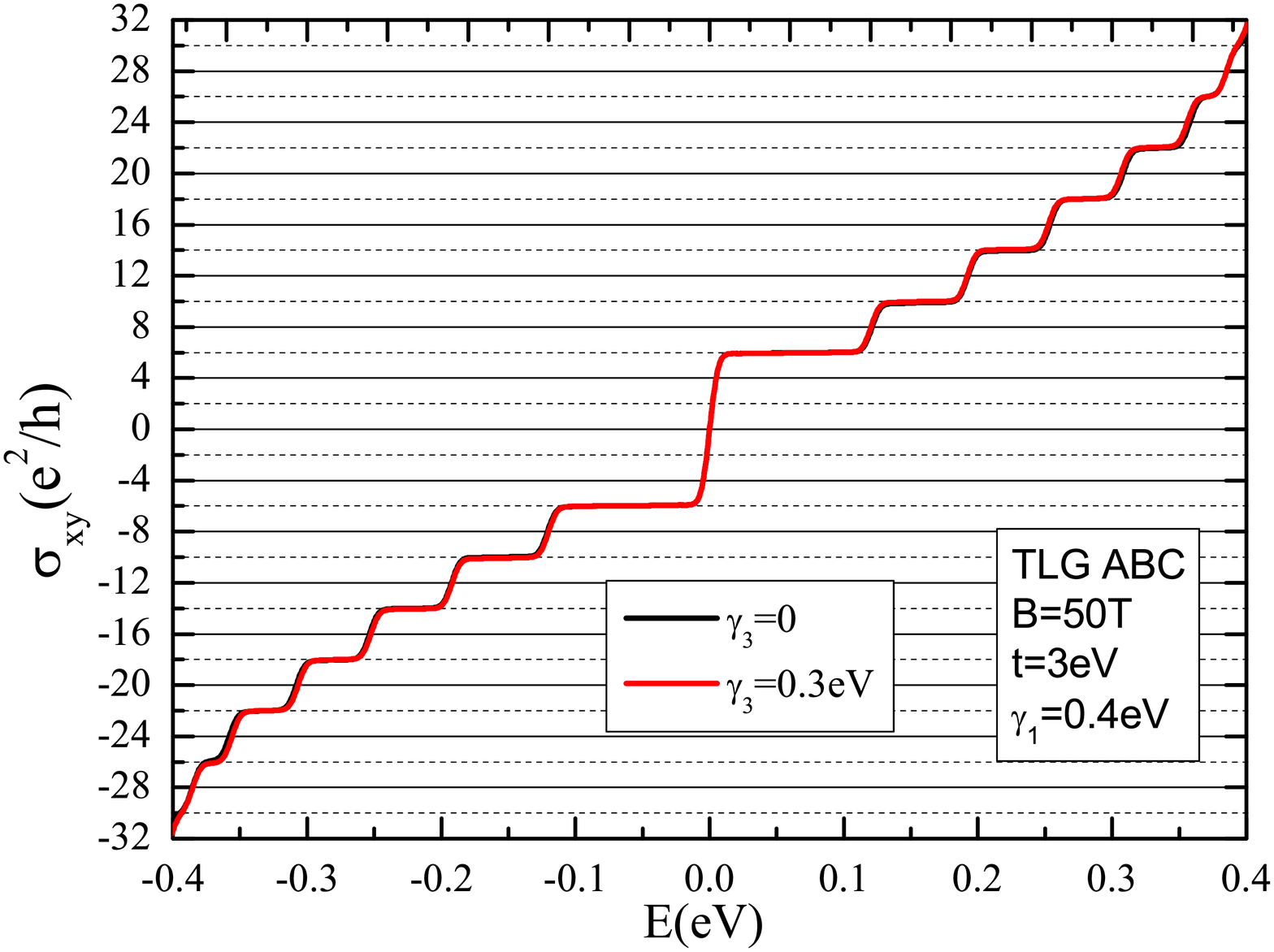}
}
\end{center}
\caption{ (Color online) Comparison of the Landau level spectrum and Hall
conductivities of ABA- and ABC-stacked trilayer graphene with (red lines) or
without (black lines) considering the interlayer hopping parameter $\protect%
\gamma _{3}$. }
\label{trilayergemma3}
\end{figure*}

In this appendix we study the effect of considering, besides $t$ and $\gamma
_{1}$, the inter-layer hopping amplitude $\gamma _{3}$ in the spectrum (see
Fig. \ref{Fig:Stacking}). In Fig. \ref{trilayergemma3}, we compare the
Landau level spectrum and Hall conductivity of ABA- and ABC-stacked TLG with
and without $\gamma _{3}$. Here we use $\gamma _{3}=0.3$~eV as it is in the
nature graphite. \cite{Partoens2006,CG07} For the considered magnetic field,
the effect of $\gamma _{3}$ in the spectrum is negligible, as seen in Fig. %
\ref{trilayergemma3}. Therefore, trigonal warping has very small effect to
the low energy spectrum of the Landau levels in the presence of high
magnetic field. In fact, this is also the case in bilayer graphene, where
the LL spectrum can be adequately described by neglecting $\gamma _{3}$ over
the field range where $l_{B}^{-1}>\frac{3}{2}a\gamma _{3}m$ (where $m\approx
0.054m_{e}$ is the effective mass in the bulk graphite). \cite{MF06} In our
calculations, the DOS and Hall conductivity are almost the same, as it is
shown in Fig. \ref{trilayergemma3}.

\bibliographystyle{apsrev4-1}
\bibliography{BibliogrGrafeno}

\end{document}